\documentclass[11pt,draftclsnofoot,onecolumn]{IEEEtran}

\usepackage{multirow}
\usepackage{CJK,indentfirst}
\usepackage[compress]{cite}
\makeatletter
\renewcommand{\citepunct}{,\penalty\@m\hskip.13emplus.1emminus.1em}
\renewcommand{\citedash}{\hbox{--}\penalty\@m}
\makeatother
\usepackage{graphicx}
\usepackage{amsmath}
\usepackage{amssymb}
\usepackage{amsfonts}
\usepackage{psfrag}
\usepackage{booktabs}
\usepackage{array}
\usepackage[named]{algo}
\usepackage{algorithmic}
\ifCLASSOPTIONcompsoc
\usepackage[tight,normalsize,sf,SF]{subfigure}
\else
\usepackage[tight,footnotesize]{subfigure}
\fi
\usepackage{color}

\IEEEoverridecommandlockouts

\renewcommand{\baselinestretch}{1.78}
\begin{document}
%\begin{CJK*}{GBK}{song}
%\CJKindent

% paper title
\title{Is Massive MIMO Energy Efficient?}

\author{\IEEEauthorblockN{Wenjia Liu, Shengqian Han and Chenyang
Yang}\thanks{The authors are with the School of Electronics and
Information Engineering, Beihang University, Beijing 100191, China (email:
\{liuwenjia, sqhan, cyyang\}@buaa.edu.cn). This work is partially supported by NSFC (China) under Grant 61120106002 and
973 Program 2012CB316003. Submission Date: Jan. 08, 2015.}}\vspace{-15mm}

%\IEEEauthorblockA{School of Electronics and Information Engineering, Beihang University, China\\
%Email: @ee.buaa.edu.cn, @ee.buaa.edu.cn,
%cyyang@buaa.edu.cn}}
%\thanks {This work was supported in part by the national key project of next generation wideband wireless communication
%networks (2011ZX03003-001), by China Postdoctoral Science Foundation (20110490271), and by the Fundamental Research Funds for the Central Universities.}

% mane the title area
\maketitle
\vspace{-5mm}
\begin{abstract}
Massive multi-input multi-output (MIMO) can support high spectral efficiency (SE) with simple linear transceivers, and is expected to provide high energy efficiency (EE). In this paper, we analyze the EE of downlink multi-cell massive MIMO systems under spatially correlated channel model, where both transmit and circuit power consumptions, training overhead, channel estimation and pilot contamination (PC) are taken into account. We obtain the maximal EE for the systems with maximum-ratio transmission (MRT) and zero-forcing beamforming (ZFBF) for given number of antennas and users by optimizing the transmit power. The closed-form expressions of approximated optimal transmit power and maximal EE, and their scaling laws with the number of antennas $M$ are derived for the systems with MRT and ZFBF. Our analysis shows that the maximal EE decreases with $M$ for both systems with and without PC, but with different descending speeds. For the system without PC, the optimal transmit power should be configured to increases with $M$, while for the system with PC, the optimal transmit power should be configured as a constant independent from $M$.  The analytical results are validated by simulations under a more realistic three-dimensional channel model.
%For a given traffic load,
%the EE-maximal massive MIMO system should be configured with the minimum number of antennas
%that can satisfy the average rate requirement.
\end{abstract}

%%%%%%%

\section{Introduction}
Massive multi-input multi-output (MIMO) has become a promising technique for  fifth-generation (5G) cellular networks. Simply by significantly increasing the number of antennas at the base station (BS), $M$, massive MIMO can support very high spectral efficiency (SE) with linear transceivers \cite{Unlimited10}. When $M \to \infty$, multi-user interference (MUI) and inter-cell interference (ICI) are randomized and averaged out without relying on sophisticated precoding and coordination, and the only limiting factor on improving SE is pilot contamination (PC) \cite{Unlimited10,ScaleUp,Com2014}.

Energy efficiency (EE) is one of critical design goal for  5G networks \cite{li2011energy}. To provide the same target data rate for a user or target sum rate for a system, MIMO systems can achieve higher EE than single antenna systems when only taking into account transmit power~\cite{EE_LTE12}. However, MIMO technique improves the SE at the cost of extra radio frequency (RF) links and complicated signal processing, both of which consume considerable circuit power \cite{EEMIMO04}. By
optimizing the configuration of spatial and frequency resources to maximize the EE of downlink multicarrier MIMO systems, it was shown in \cite{XZK2013OFDM} that the EE increases with the SE only when the frequency resource is still available, and the maximal EE of traditional MIMO system reduces with the increase of the number of antennas.

Massive MIMO systems are expected to be energy efficient, since the transmit power can be significantly reduced owing to the large array gain and multi-user multiplexing gain  \cite{EESEuplink}. Moreover, by installing a large number of antennas and with the low transmit power per-antenna, inexpensive components can be used to build the system~\cite{Com2014}.
In the past several years, valuable research results have been obtained for analyzing the EE of massive MIMO
systems. For the uplink, it was shown in  \cite{EESEuplink} that massive MIMO has a large potential in improving EE by studying the power scaling law, where the EE is defined as the sum rate divided by the transmit power. The results show that to achieve the same sum rate as a single antenna system, when $M \to \infty$, the transmit power of massive MIMO system reduces with the law of $1/M$ if the BS has perfect channel information and with $1/\sqrt{M}$ if the BS has estimated channels with PC.
When the circuit power consumption is  considered, however, a massive MIMO system is more energy efficient than traditional MIMO systems only when the average channel
gain is small or the circuit power consumption is low \cite{MassiveEE_TCOM14}. For the downlink, it was shown in
\cite{WCNC14} that the optimal transmit power of BS needs to increase with $M$ in order to
maximize the EE of a massive MIMO systems with zero-forcing beamforming (ZFBF), with given numbers of antennas
and users. The SE-EE relationship of massive MIMO systems was investigated in~\cite{ICC14}, which showed that the
EE will increase exponentially with a linear SE loss by reducing the circuit power. This result can also be
interpreted as: \emph{massive MIMO is not energy efficient to achieve very high SE}, because a linear increase of SE leads to an exponential decrease of EE. While existing results provide useful insight on understanding the potential EE gain of massive MIMO, and partially answer the question of whether massive MIMO is
energy efficient, a quantitative characterization of how the EE scales with $M$ remains open. Moreover, the prior results are all obtained under the following assumptions: (1)
single cell system without ICI, (2) perfect channel information or channel estimation without PC, and (3) independent and identically distributed
(i.i.d.) channels.

In this paper, we investigate the EE of downlink multi-cell massive MIMO systems and strive to answer the following questions. (1) What is the maximal EE of the massive MIMO systems with the given number of antennas? (2) How should the massive MIMO systems be configured in practice to achieve the maximal EE? (3) What is the impact of PC, channel estimation errors, multi-cell setting, channel correlation, and
specific beamforming?

In real-world cellular networks, the traffic load in a cell may vary with time (e.g., day and night) and location (e.g., urban and suburban). This implies that in general the number of users is not a parameter that can be configured. Therefore, we optimize the transmit power of the BS to maximize the average EE of the system, given maximal ratio transmission (MRT) or ZFBF together with equal power allocation among multiple users, which are widely accepted precoders for massive MIMO~\cite{EESEuplink,ScaleUp}. To answer the proposed questions, we derive and analyze the closed-form expressions of the optimal transmit power and maximal average EE and their scaling laws with $M$. Since PC is the limiting factor for massive MIMO systems, which however can be mitigated by some recently-proposed pilot decontamination methods such as  \cite{CZL_GlobalSIP14}, we consider both the systems  with and without PC. The analytical results are validated by simulations  with a more realistic channel model, where different-level power consumption parameters, ranging from those in existing systems to those that might be possible in the future, are taken into account.

The rest of the paper is organized as follows. In section II, we provide the system and power consumption models. In section III, we derive the closed-form expressions of the average data rates, respectively for the massive MIMO systems with MRT and ZFBF. In section IV, we find the maximal EE by optimizing the transmit power and derive the scaling laws of the EE and optimal transmit power with respective to the number of antennas. In section V, we validate the analytical results by simulations. Finally, we conclude the paper.

\section{System Model and Power Consumption Model}
\subsection{System Model}
Consider a downlink massive MIMO system consisting of $L$ non-coordinated cells, where each BS equipped with $M$ transmit antennas serves $K$ single-antenna users either with MRT or with ZFBF. The transmit power of the BS is $P$, which is assumed to be equally allocated to multiple users. We assume block fading channel, where the channels are constant in a time-frequency coherence block with $T$ channel uses and independent among blocks. The channel from the $l$-th BS (denoted by $\text{BS}_l$) to the $k$-th user in the $j$-th cell (denoted by $\text{UE}_{jk}$) is modeled~as\cite{HowManyAntennas13}
\begin{equation}\label{E:Channel_Model}
   \mathbf{h}_{ljk}=\sqrt{\alpha\chi_{lj} \rho}\ \tilde{\mathbf{R}}\tilde{\mathbf{h}}_{ljk}\in \mathbb{C}^{M\times 1},
\end{equation}
where $\alpha$ is the average channel gain including path loss and shadowing, $\chi_{lj} = 1$ if $l=j$ and $\chi_{lj} = \chi$ if $l\neq j$ with $\chi\in (0,1]$ reflecting the difference of channel gains from the serving BS and interfering BSs, $\tilde{\mathbf{R}}\in \mathbb{C}^{M\times N}$ is composed of $N$ columns of a unitary $M\times M$ matrix with $N\leq M$, which models the correlated channels with $N$ angles of arrival, $\rho=\frac{M}{N}$, $\tilde{\mathbf{h}}_{ljk}\sim\mathcal{CN}(\mathbf{0},\mathbf{I}_N)$ is the i.i.d. channel vector, $\mathcal{CN} (\mathbf{0},\mathbf{I}_N)$ denotes zero-mean complex Gaussian distribution with covariance matrix $\mathbf{I}_N$, and $\mathbf{I}_N$ is the $N$-dimensional identity matrix. The channel correlation matrix is $\mathbf{R}_{lj}\triangleq\chi_{lj} \alpha\rho\tilde{\mathbf{R}}\tilde{\mathbf{R}}^H\in \mathbb{C}^{M\times M}$.\footnote{In order to obtain simple form expressions that can reflect the inherent nature of the system, we consider that all users have the same correlation matrix as in \cite{HowManyAntennas13}. Note that the channels among users are independent, but their instantaneous channel vectors are less orthogonal  due to the same correlation matrix. This leads to the statistically largest MUI and ICI and therefore reflects the worst-case performance of massive MIMO systems. Later on, we will use simulation under more realistic spatial correlated channel model to valid the analytical results derived from the model in \eqref{E:Channel_Model}.}

We consider a time-division duplex system, where the channels can be estimated with the pilot sequences sent by users. During the uplink training phase, $K$ users in the same cell transmit orthogonal pilot sequences each with $T_{\rm{tr}}$ channel uses and transmit power $P_{\text{tr}}$. The pilots received at $\text{BS}_j$ will be contaminated  if the same set of pilot sequences used in the $j$-th cell are reused in other cells and the traditional training signal and channel estimation methods are applied. Nonetheless, if some pilot decontamination methods such as that in \cite{CZL_GlobalSIP14} can be applied, the PC will be largely mitigated without increasing the training overhead. Therefore, we consider two extreme cases for the systems with and without PC to simplify the analysis.

Denote $\mathcal{S}_j$ as the index set of the cells using the same set of pilot sequences as the $j$-th cell, and $L_{Pj}=|\mathcal{S}_j|$ as the number of these cells, where
$\text{UE}_{lk}$ transmits the same pilot sequence as $\text{UE}_{jk}$ for $l \in \mathcal{S}_j$.
For simplicity, we assume that $L_{Pj}=L_P$, $\forall j$. Then: (1) \emph{for the system with PC}, $\mathcal{S}_j\!=\!\{1,2,\ldots,L\}$ and $L_{P}\!=\!L$, where the same set of orthogonal pilot sequences are reused in all the $L$ cells; (2) \emph{for the system without PC}, we set $\mathcal{S}_j=\{j\}$ and $L_{P}=1$.

The minimum mean-square error (MMSE)  channel estimate can be obtained at  BS$_j$ as
\begin{equation}\label{E:MMSE_CHest}
\hat{\mathbf{h}}_{jjk}= \mathbf{R}_{jj}\mathbf{Q}\left(\mathbf{h}_{jjk} +\sum_{l\in \mathcal{S}_j\backslash j}\mathbf{h}_{jlk}+\frac{1}{\sqrt{KT_{\text{tr}}P_{\text{tr}}}} \mathbf{n}_{jk}^{\text{tr}}\right),
\end{equation}
where $\mathbf{Q}\!=\!\left((1\!+\!\chi(L_P\!-\!1))\alpha\rho  \tilde{\mathbf{R}}\tilde{\mathbf{R}}^H
\!+\!\frac{1}{\gamma_{\text{tr}}}\mathbf{I}_M\right)^{-1}$,  $\mathbf{n}_{jk}^{\text{tr}}\sim
\mathcal{CN}(\mathbf{0},\sigma^2_{\rm{tr}}\mathbf{I}_M)$ is the noise at the BS, and $l\in \mathcal{S}_j\backslash j$ denotes the index set of all interfering cells using the same
set of orthogonal pilot sequences as the $j$-th cell. We define $\gamma_{\text{tr}}\triangleq
\frac{K T_{\rm{tr}} P_{\text{tr}}}{\sigma^2_{\rm{tr}}}$ to reflect the signal to noise ratio (SNR) of the uplink
training.

It is not hard to find that $\hat{\mathbf{h}}_{jjk}\sim \mathcal{CN}(\mathbf{0},\mathbf{\Phi}_{jj})$, where $\mathbf{\Phi}_{jj}\triangleq\mathbf{R}_{jj}\mathbf{Q}\mathbf{R}_{jj}= v\alpha\rho \tilde{\mathbf{R}}\tilde{\mathbf{R}}^H$, and $v \triangleq \frac{\alpha\rho}{(1+\chi(L_P-1))\alpha\rho+\frac{1}{\gamma_{\rm{tr}}}}$. The parameter $v$ reflects the channel estimation accuracy. Specifically, under more correlated channel (i.e., with larger $\rho$), with higher uplink training SNR, without PC (i.e., $L_P =1$), with larger average channel gain $\alpha$, and with weaker ICI (i.e., with smaller $\chi$), the value of $v$ is larger, indicating a more accurate channel estimate. Considering the orthogonality between the MMSE channel estimate and estimation errors, we have $\mathbf{h}_{jjk}=\hat{\mathbf{h}}_{jjk}+\check{\mathbf{h}}_{jjk}$, where $\check{\mathbf{h}}_{jjk}\sim \mathcal{CN}(\mathbf{0},\mathbf{R}_{jj}-\mathbf{\Phi}_{jj})$ is the channel estimation error \cite{MMSE_book}.

The  downlink signal received at $\text{UE}_{jk}$ is given~by
\begin{equation}\label{E:ReceiveSignal}
y_{jk} = \sqrt{\frac{P}{K}}\sum_{l=1}^L\sum_{m=1}^{K} \mathbf{h}_{ljk}^H\sqrt{\lambda_{lm}}\mathbf{w}_{lm} x_{lm}+n_{jk},
\end{equation}
where $\mathbf{w}_{lm}\in \mathbb{C}^{M\times 1}$ is the beamforming vector of $\text{BS}_l$ to $\text{UE}_{lm}$, $\lambda_{lm}=\frac{1}{\|\mathbf{w}_{lm}\|^2}$, $x_{lm}$ is the transmit
signal with $\mathbb{E}\{|x_{lm}|^2\}=1$, $n_{jk}\sim \mathcal{CN}(0,\sigma^2)$ is the noise at the user, $\|\cdot\|$ denotes the Euclidean norm, and $\mathbb{E}\{\cdot\}$
denotes the expectation. For MRT, $\mathbf{w}_{lm}=\hat{\mathbf{h}}_{llm}$. For ZFBF, $\mathbf{w}_{lm}$ is the $m$-th column of $\hat{\mathbf{H}}_{ll}(\hat{\mathbf{H}}_{ll}^H \hat{\mathbf{H}}_{ll})^{-1}$, and $\hat{\mathbf{H}_{ll}}=[\hat{\mathbf{h}}_{ll1},\ldots, \hat{\mathbf{h}}_{llK}]$ is the estimated channel matrix at BS$_l$ for all the $K$ users in the $l$-th cell.

Every user experiences \emph{MUI} (when using ZFBF, it comes from the channel estimation errors), \emph{coherent ICI} caused by PC, and \emph{non-coherent ICI} generated by the signals from the interfering BSs. From \eqref{E:ReceiveSignal}, the signal to interference plus noise ratio (SINR) of $\text{UE}_{jk}$ can be expressed~as
\begin{equation}\label{E:SINR}
\gamma_{jk}=\frac{\lambda_{jk}\|\mathbf{h}_{jjk}^H\mathbf{w}_{jk}\|^2} {\underbrace{\sum_{l\in\mathcal{S}_j\backslash j}\lambda_{lk}\|\mathbf{h}_{ljk}^H\mathbf{w}_{lk}\|^2}_{\text{coherent ICI caused by PC}} + \underbrace{\sum_{\{l\!\in\! \mathcal{S}_j,m\!\neq\! k\}\!\cup\!\{l\!\notin \!\mathcal{S}_j,m\}}\lambda_{lm} \|\mathbf{h}_{ljk}^H\mathbf{w}_{lm}\|^2}_{\text{MUI and non-coherent ICI}}+\frac{K\sigma^2}{P}},
\end{equation}
where $\{l\!\in\! \mathcal{S}_j,m\!\neq\! k\}\!\cup\!\{l\!\notin \!\mathcal{S}_j,m\}$ denotes the set of users transmitting different pilot sequences with $\text{UE}_{jk}$.

Then, the average sum data rate of the $j$-th cell with transmit power $P$ at  BS$_j$ can be obtained~as
\begin{equation}\label{E:Average_UserRate}
R_{j}(P)=B\sum_{k=1}^K\mathbb{E}_{\mathbf{h}}
\left\{\log_2\left(1+\gamma_{jk}\right)\right\}\triangleq \sum_{k=1}^{K}R_{jk}(P),
\end{equation}
where $B$ is the system bandwidth, and $R_{jk}(P)$ is the average data rate of $\text{UE}_{jk}$.

\subsection{Power Consumption Model}
The power consumed for downlink transmission by a BS and the circuit powers for operating the BS in transmitting
and receiving phases can be modeled~as~\cite{E3F_EARTH,GreenTouch_Power}\footnote{A appropriate power consumption model is critical for evaluating the EE of a system. According to the analysis in
\cite{GreenTouch_Power}, the circuit power consumption of a massive MIMO BS can be modeled in the same way as a
traditional BS, but the corresponding parameters will differ. Note that in (\ref{E:PowerModel2}) the feeder loss
is removed considering that  massive MIMO systems will employ active antennas where the RF module is integrated
into antenna, which is different from a traditional BS with passive antennas~\cite{EETrans_LTE13}. }
\begin{equation}\label{E:PowerModel2}
P_{\text{BS}}=\frac{(1\!-\!\frac{KT_{\rm{tr}}}{T}) \frac{P}{\eta_{\rm{PA}}}\!+\!(1\!-\!\frac{KT_{\rm{tr}}}{T})P_{\rm{BB}_2} \!+\!\frac{KT_{\rm{tr}}}{T}P_{\rm{CE}}\!+\!M(P_{\rm{RF}}\!+\! (1\!-\!\frac{KT_{\rm{tr}}}{T})P_{\rm{BB}_{1d}}\!+\!\frac{KT_{\rm{tr}}}{T}P_{\rm{BB}_{1u}})} {(1-\sigma_{\rm{DC}})(1-\sigma_{\rm{MS}})(1-\sigma_{\rm{cool}})},
\end{equation}
where $\eta_{\rm{PA}}$ is the power amplifier efficiency, $\sigma_{\rm{DC}}$, $\sigma_{\rm{MS}}$ and
$\sigma_{\rm{cool}}$ are respectively the loss factors of direct-current to direct-current power supply, main
supply and cooling \cite{E3F_EARTH}, $P_{\rm{BB}_2}$ is the baseband signal processing power consumed for
computing beamforming vectors, $P_{\rm{CE}}$ is the signal processing power consumed for channel estimation, $P_{\rm{RF}}$ is the RF power consumption, and $P_{\rm{BB}_{1d}}$ and $P_{\rm{BB}_{1u}}$ are respectively other baseband processing power consumption in the downlink transmission phase and in the uplink training phase. According to
\cite{EELSAS_Marzetta13}, $P_{\rm{BB}_2}$ can be modeled as
$P_{\rm{BB}_2}=\frac{M(K+\delta_{ZF}K^2)R_{flops,0}}{\eta_C}$, where $R_{flops,0}$ is the floating-point
operations per-second (flops) per-antenna for each user, $\eta_C$ is the power efficiency of computing measured
in flops/W, and we use a binary variable $\delta_{ZF}$ to differentiate MRT ($\delta_{ZF}=0$) and ZFBF
($\delta_{ZF}=1$) in computational complexity. $P_{\rm{CE}}$ can be modeled as $P_{\rm{CE}}=\frac{M\log_2(KT_{\rm{tr}})R_{flops,0}}{\eta_C}$.

In order to differentiate the impact of transmit and circuit power consumptions on EE, we define an \emph{equivalent circuit power consumed at each antenna except beamforming} as $P_c\triangleq
\frac{P_{\rm{RF}}+(1\!-\!\frac{KT_{\rm{tr}}}{T})P_{\rm{BB}_{1d}} +\frac{KT_{\rm{tr}}}{T}P_{\rm{BB}_{1u}}}{(1-\sigma_{\rm{DC}})(1-\sigma_{\rm{MS}}) (1-\sigma_{\rm{cool}})}$, and
an \emph{equivalent circuit power consumed  for beamforming at each antenna for each user}~as
$P_{sp}\triangleq \frac{R_{flops,0}}{\eta_C (1-\sigma_{\text{DC}})(1-\sigma_{\text{MS}})(1-\sigma_{\text{cool}})}$. Then, (\ref{E:PowerModel2}) can be rewritten~as
\begin{align}\label{E:PowerModel}
P_{\text{BS}}
&= \underbrace{(1-\tfrac{KT_{\rm{tr}}}{T})\eta P}_{\text{transmit power}}+\underbrace{M P_{c}+M((1\!-\!\tfrac{KT_{\rm{tr}}}{T})(K+\delta_{ZF}K^2) +\tfrac{KT_{\rm{tr}}}{T}\log_2(KT_{\rm{tr}})) P_{sp}}_{\text{circuit power}}\nonumber\\
&\approx (1-\tfrac{KT_{\rm{tr}}}{T})\eta P+MP_0,
\end{align}
where $\eta\triangleq
\frac{1}{\eta_{\rm{PA}} (1-\sigma_{\rm{DC}})(1-\sigma_{\rm{MS}})(1-\sigma_{\rm{cool}})}$, $P_0=\frac{P_{\rm RF}+P_{\rm{BB}_{1d}}}{(1-\sigma_{\text{DC}})(1-\sigma_{\text{MS}})(1-\sigma_{\text{cool}})} +(K+\delta_{ZF}K^2)P_{sp}$ is an \emph{equivalent circuit power consumption per-antenna}, and the approximation comes from the similarity of the signal processing power consumed in uplink training and downlink transmission~\cite{GreenTouch_Power}.\vspace{-3mm}

\subsection{Downlink EE}

The downlink EE is defined as the ratio of the average downlink throughput to the total power consumption of the $L$ BSs, where the throughput is the sum rate of all cells excluding the uplink training overhead. From \eqref{E:Average_UserRate} and \eqref{E:PowerModel}, we can express the downlink EE of the network as
\begin{equation}\label{E:AreaEE}
\mathrm{EE}(P)=\frac{(1-\frac{KT_{\rm{tr}}}{T})\cdot \sum_{j=1}^LR_{j}(P)}{L((1-\frac{KT_{\rm{tr}}}{T})\eta P+MP_0)}.
\end{equation}

\section{Average Data Rate Analysis}
To obtain a closed-form expression of $\mathrm{EE}(P)$ for optimization, we derive the asymptotic data rate for large system, where $M$ and $K$ grow infinitely while ${M}/{K}$ is finite. According to the random-matrix theory \cite{Book_RandomMatrix}, the asymptotic rate converges in mean square to the average rate. Hence, we can use the asymptotic data rate as the average rate.\vspace{-2mm}

\subsection{Average Data Rate for MRT}
For the system using MRT, $\mathbf{w}_{jk}=\hat{\mathbf{h}}_{jjk}$ and the SINR of $\text{UE}_{jk}$ in \eqref{E:SINR} can be rewritten~as
\begin{equation}\label{E:Inst_SINR_MRT}
\gamma_{jk}^{\rm{M}}=\frac{\lambda_{jk}\|\mathbf{h}_{jjk}^H \mathbf{\hat{h}}_{jjk}\|^2} {\underbrace{\sum\limits_{l\in \mathcal{S}_j\backslash j}\lambda_{lk}\|\mathbf{h}_{ljk}^H\mathbf{\hat{h}}_{llk}\|^2}_{\text{coherent ICI caused by PC}} +\underbrace{\sum\limits_{\{l\in \mathcal{S}_j,m\neq k\}\cup \{l\notin \mathcal{S}_j,m\}}\lambda_{lm}\|\mathbf{h}_{ljk}^H \mathbf{\hat{h}}_{llm}\|^2}_{\text{MUI and non-coherent ICI}} +\frac{K\sigma^2}{P}}.
\end{equation}

By using the methods of asymptotic analysis in \cite{HowManyAntennas13} but reserving the terms related to channel estimation errors, we can derive the asymptotic data rate provided by each  BS with MRT as
\begin{align}
\bar{R}_{\rm{BS}}^{\rm{M}}\!&
=\!BK\log_2\left(\!1\!+\!\frac{S^{\rm{M}}} {I_{P}^{\rm{M}}+I_{nP}^{\rm{M}}+\!\frac{K\sigma^2}{v\alpha P}}\!\right),\label{E:Asymp_Rate_CH2_MRT}
\end{align}
where $v=\frac{\alpha\rho}{\alpha\rho\bar{L}_P+\frac{1}{\gamma_{\rm{tr}}}}\leq 1$ reflects the accuracy of the estimated channel,  and
\begin{align}
S^{\rm{M}}&\triangleq M+\tfrac{1}{\gamma_{\rm{tr}}\alpha}+(\bar{L}_P-1)\rho, \label{E:Asymp_S_CH2_MRT}\\
I_{P}^{\rm{M}}&\triangleq\chi(\bar{L}_P-1)(M-\rho), \quad \quad
I_{nP}^{\rm{M}}\triangleq (K\bar{L}-1)(\rho\bar{L}_P+\tfrac{1}{\gamma_{\rm{tr}}\alpha}), \label{E:Asymp_ICI_noPC_MRT}
\end{align}
are respectively the average receive powers of the desired signal, coherent ICI, MUI and non-coherent ICI, all normalized by $v\alpha$, $\bar{L}_P \triangleq \sum_{l\in \mathcal{S}_j}\chi_{lj}= 1+\chi(L_P-1)$ and $\bar{L} \triangleq \sum_{l=1}^{L}\chi_{lj}=1+\chi(L-1)$.
Both the average signal power $v\alpha S^{\rm{M}}=\alpha((M-\rho)v+\rho)$ and the coherent ICI power $v\alpha I_{P}^{\rm{M}}=\alpha v\chi(\bar{L}_P-1)(M-\rho)$ increase with $v$, which means that they become higher when the channel estimate becomes accurate, and the non-coherent ICI power $v\alpha I_{nP}^{\rm{M}}=\alpha (K\bar{L}-1)\rho$ does not depend on the channel estimation accuracy.\vspace{-3mm}

%It can be observed that for the system with PC, i.e., $\bar{L}_P > 1$, both $S^{\rm{M}}$ and $I_P^{\rm{M}}$ increase with $M$. For small channel gain $\alpha$ and moderate $M$ such that $I_{nP}^{\rm{M}}\gg I_{P}^{\rm{M}}$, the system will be limited by non-coherent ICI. From $I_{P}^{\rm{M}}> I_{nP}^{\rm{M}}$, we can see that only for large $\alpha$ (say, $\alpha >\frac{1}{\gamma_{\rm{tr}}(\frac{\chi(\bar{L}_P-1)(M-\rho)}{K\bar{L}-1} -\rho\bar{L}_P)}$) and very large $M$ ($M >\rho+\frac{(K\bar{L}-1)(\rho\bar{L}_P+\frac{1}{\gamma_{\rm{tr}}\alpha})} {\chi(\bar{L}_P-1)}$), the system will be limited by PC.
%, as analyzed in \cite{ScaleUp,HowManyAntennas13}.

\subsection{Average Data Rate for ZFBF}
For the system using ZFBF, the beamforming matrix of $\text{BS}_j$ is $\mathbf{W}_j=\mathbf{\hat{H}}_{jj}\left(\mathbf{\hat{H}}_{jj}^H\mathbf{\hat{H}}_{jj}\right)^{-1}$ and the beamforming vector for $\text{UE}_{jk}$ can be expressed as $\mathbf{w}_{jk}\!=\! \mathbf{\Pi}_{j}^{\rm{Z}}\mathbf{\hat{h}}_{jjk}$, where $\mathbf{\Pi}_j^{\rm{Z}}\!=\!\mathbf{\hat{H}}_{jj} \left(\!\mathbf{\hat{H}}_{jj}^H\mathbf{\hat{H}}_{jj}\!\right)^{-2} \!\mathbf{\hat{H}}_{jj}^H$ \cite{LargeSystem}. From \eqref{E:SINR}, the instantaneous SINR of $\text{UE}_{jk}$ with ZFBF can be obtained~as
\begin{equation}\label{E:Inst_SINR_ZFBF}
\gamma_{jk}^{\rm{Z}}=\frac{\lambda_{jk}\|\mathbf{h}_{jjk}^H \mathbf{\Pi}_{j}^{\rm{Z}}\mathbf{\hat{h}}_{jjk}\|^2} {\sum\limits_{l\in \mathcal{S}_j\backslash j}\lambda_{lk}\|\mathbf{h}_{ljk}^H\mathbf{\Pi}_{l}^{\rm{Z}}\mathbf{\hat{h}}_{llk}\|^2 +\sum\limits_{\{l\in \mathcal{S}_j,m\neq k\}\cup \{l\notin \mathcal{S}_j,m\}}\lambda_{lm}\|\mathbf{h}_{ljk}^H\mathbf{\Pi}_{l}^{\rm{Z}}\mathbf{\hat{h}}_{llm}\|^2 +\frac{K\sigma^2}{P}}.
\end{equation}
The asymptotic data rate provided by each BS with ZFBF can be derived as (see Appendix \ref{P:AverageRate_ZF})
\begin{align}
\bar{R}_{\text{BS}}^{\rm{Z}}&=
\!BK\log_2\left(\!1\!+\!\frac{S^{\rm{Z}}} {I_{P}^{\rm{Z}}+I_{nP}^{\rm{Z}}+\!\frac{K\sigma^2}{v\alpha P}}\!\right),\label{E:Asymp_Rate_CH2_ZFBF}
\end{align}
where the average receive powers of the desired signal, coherent ICI and the sum power of MUI and non-coherent ICI normalized by $v \alpha$ are respectively
\begin{align}
S^{\rm{Z}}&\triangleq M+\tfrac{1}{\gamma_{\rm{tr}}\alpha}+(\bar{L}_P-1-K)\rho, \label{E:Asymp_S_CH2_ZF}\\
I_P^{\rm{Z}}&\triangleq \chi(\bar{L}_P-1)(M-2\rho K), \quad\quad
I_{nP}^{\rm{Z}} \triangleq (K\bar{L}-1)(\rho\bar{L}_P+\tfrac{1}{\gamma_{\rm{tr}}\alpha})
-\rho(K-1).\label{E:Asymp_ICI_noPC_ZF}
\end{align}
Similar with MRT, both the signal and coherent ICI powers increase with $v$. However, different from MRT, the sum of the noncoherent ICI and MUI power $v\alpha I_{nP}^{\rm{Z}}=\alpha(K\bar{L}-1)\rho-v\alpha \rho(K-1)$ decreases with $v$, since accurate channel estimate induces less MUI.

\emph{Remark 1 (Maximum data rate for given $P$ and infinite $M$)}: By comparing the asymptotic rates of MRT and ZFBF, we can find that for arbitrary system configuration, $S^{\rm{M}}\!>\!S^{\rm{Z}}$, $I_{P}^{\rm{M}}\!>\!I_P^{\rm{Z}}$ and $I_{nP}^{\rm{M}}\!>\!I_{nP}^{\rm{Z}}$. Consequently, $\lim_{M\rightarrow \infty}S^{\rm{M}}\!=\!\lim_{M\rightarrow \infty}S^{\rm{Z}}\!=\!M$, but $I_{nP}^{\rm{M}}\!>\!I_{nP}^{\rm{Z}}$ still holds that is independent from $M$. For the system without PC (i.e., $\bar{L}_P=1$), we have $I_{P}^{\rm{Z}}\!=\!I_{P}^{\rm{M}}\!=\!0$, and hence the data rate achieved by ZFBF is higher than MRT, where the rate gap approaches to a constant, which is $\lim_{M\rightarrow \infty} (\bar{R}_{\text{BS}}^{\rm{Z}}-\bar{R}_{\text{BS}}^{\rm{M}}) \!=\!BK\log_2\left(1\!+\!\frac{I_{nP}^{\rm{M}}-I_{nP}^{\rm{Z}}} {I_{nP}^{\rm{Z}}+\frac{K\sigma^2}{v\alpha P}}\right)$. For the system with PC (i.e., $\bar{L}_P > 1$), we can obtain $\lim_{M\rightarrow \infty}(I_{P}^{\rm{M}}\!+\!I_{nP}^{\rm{M}})\!=\!\lim_{M\rightarrow \infty}(I_P^{\rm{Z}}\!+\!I_{nP}^{\rm{Z}})\!=\!\chi(\bar{L}_P\!-\!1)M$, which indicates the same data rate for MRT and ZFBF by recalling that $\lim_{M\rightarrow \infty}S^{\rm{M}}\!=\!\lim_{M\rightarrow \infty}S^{\rm{Z}}$.

\emph{Remark 2 (Maximum data rate for given $M$ and infinite $P$)}: From \eqref{E:Asymp_Rate_CH2_MRT} and \eqref{E:Asymp_Rate_CH2_ZFBF}, we have $\lim_{P\rightarrow \infty} \bar{R}_{\text{BS}}^{\rm{M}}=BK\log_2\left(1+\frac{S^{\rm{M}}} {I^{\rm{M}}}\right)$ and $\lim_{P\rightarrow \infty}\bar{R}_{\text{BS}}^{\rm{Z}}=BK\log_2\left(1+ \frac{S^{\rm{Z}}}{I^{\rm{Z}}}\right)$ for MRT and ZFBF, respectively, where $I^{\rm M}=I_P^{\rm{M}} +\!I_{nP}^{\rm{M}}$ and $I^{\rm Z}= I_P^{\rm{Z}} +\!I_{nP}^{\rm{Z}}$. For the single-cell systems with perfect channel estimation (i.e., $\gamma_{\rm{tr}} \to \infty$)  considered in \cite{WCNC14, ICC14}, we can obtain that $I^{\rm{M}}=(K-1)\rho$ due to MUI for MRT and $I^{\rm{Z}}=0$ for ZFBF, which indicates that with MRT the system is interference-limited for large $P$ and achieves a finite maximum data rate, but with ZFBF the rate of the system increases with $P$ without limit. However, for the multi-cell scenario with channel estimation errors, the results are different. First, we have $I^{\rm{M}}>0$ and $I^{\rm{Z}}>0$ in multi-cell systems, because $L>1$ and $\bar{L}>1$. Second, imperfect channel estimation will lead to $I^{\rm{M}}>0$ and $I^{\rm{Z}}>0$ even for single-cell systems. Therefore, \emph{once} ICI \emph{or} channel estimation errors is considered, both systems with MRT and ZFBF become interference-limited, whose maximum achievable rates are finite even when $P \to \infty$.\vspace{-0mm}

\section{Maximal Energy Efficiency Analysis}
In this section, we optimize the transmit power to maximize EE, respectively for the massive MIMO systems with MRT and ZFBF. We then analyze how to configure the transmit power to maximize the EE for given numbers of antennas and users and the resulting maximal EE.
To show the potential EE gain of massive MIMO, we do not consider the transmit power constraint. Based on \eqref{E:AreaEE}, \eqref{E:Asymp_Rate_CH2_MRT} and \eqref{E:Asymp_Rate_CH2_ZFBF}, the maximal EE can
be found from the following problem
\begin{align}\label{E:OptimizeEE}
\underset{P}{\max}&\ \mathrm{EE}(P)=\frac{(1-\frac{KT_{tr}}{T})BK\log_2\left(1+\frac{SP} {IP+G}\right)}{(1-\frac{KT_{tr}}{T})\eta P+MP_0} \\
{s.t.}& \quad P \geqslant 0,\nonumber
\end{align}
where $S=S^{\rm{M}}$ in \eqref{E:Asymp_S_CH2_MRT} and $I=I_P^{\rm{M}}\!+\!I_{nP}^{\rm{M}}$ in \eqref{E:Asymp_ICI_noPC_MRT} for MRT, $S=S^{\rm{Z}}$ in \eqref{E:Asymp_S_CH2_ZF} and $I\!\!=\!\!I_P^{\rm{Z}}\!+\!I_{nP}^{\rm{Z}}$ in  \eqref{E:Asymp_ICI_noPC_ZF}  for ZFBF, and $G=\frac{K\sigma^2}{v\alpha}$ for both.

It is not hard to show that the numerator of $\mathrm{EE}(P)$ in (\ref{E:OptimizeEE}) is concave, the denominator is convex, and both are differential. Hence, the EE is a pseudoconcave function with respect to $P$. This suggests that the problem has a globally optimal solution, which can be found from
the \emph{Karush-Kuhn-Tucker} (KKT) condition.

If the circuit power consumption
$MP_0=0$, it is not hard to obtain from the KKT condition that the optimal transmit power $P^*=0$, then the corresponding data rate is zero. The
maximal EE can be obtained by L'Hospital's rule  as
\begin{align}\label{E:MaximalEE_noPc_MRT}
\mathrm{EE}^*=\frac{BK}{(1-\frac{KT_{tr}}{T})\eta} \lim_{P\rightarrow
0}\frac{\log_2\left(1+\frac{SP}{IP+G}\right)}{P} =\frac{BK\log_2e}{(1-\frac{KT_{tr}}{T})\eta}\cdot \frac{S}{G}.
\end{align}

If the circuit power consumption $MP_0>0$, which is true in practice, after some regular manipulations we can obtain the KKT condition as
\begin{equation}\label{E:KKT_Accurate}
\left(P^*+\frac{MP_0}{(1-\frac{KT_{\rm{tr}}}{T})\eta}\right)\frac{SG} {((S+I)P^*+G)(IP^*+G)} -\ln\left(1+\frac{SP^{*}}{IP^{*}+G}\right)=0.
\end{equation}

\subsection{Optimal Transmit Power Analysis}

By expressing $\frac{MP_0}{(1-\frac{KT_{\rm{tr}}}{T})\eta}$ as a function as $P^*$ from \eqref{E:KKT_Accurate}, it is easy to show that its first-order derivative over $P^*$ is positive, which gives rise to the following proposition.

\textbf{Proposition 1:} $P^*$ increases monotonically with $\frac{MP_0}{(1-\frac{KT_{\rm{tr}}}{T})\eta}$.

That is to say, in order to maximize the EE of the massive MIMO system the transmit power should increase with the number of antennas $M$, equivalent circuit power consumption per-antenna $P_0$, training overhead $KT_{\rm{tr}}$, and equivalent power amplifier efficiency $\frac{1}{\eta}$.

To understand the behavior of massive MIMO systems with given number of antennas, we derive the scaling law of the optimal transmit power and maximal EE with respect to $M$. To this end, we need to find the closed-form expression of $P^*$ from the transcendental equation in (\ref{E:KKT_Accurate}), which is very difficult if not possible. To tackle the difficulty, we introduce some approximations, which are accurate for massive MIMO systems. With the derivations in Appendix \ref{P:Optimal_P_MRT}, the approximate optimal transmit power can be obtained~as
\begin{equation}\label{E:Optimal_P}
P^*\approx\sqrt{\frac{MP_0K\sigma^2(\frac{1}{\rho \gamma_{\rm{tr}}\alpha}+\bar{L}_P)}{(1-\frac{KT_{\rm{tr}}}{T})\eta \alpha}} \sqrt{\frac{\frac{1}{I}-\frac{1}{S+I}} {\ln\left(1+\frac{S}{I}\right)}},
\end{equation}
which is accurate for large value of $M$.
%Because  $S\approx M$ for large $M$ as shown in \eqref{E:Asymp_S_CH2_MRT} and \eqref{E:Asymp_S_CH2_ZF}, the optimal transmit power can be approximated as $P^*\approx \sqrt{\frac{P_0K\sigma^2(\frac{1}{\rho \gamma_{\rm{tr}}\alpha}+\bar{L}_P)}{(1-\frac{KT_{\rm{tr}}}{T})\eta \alpha}} \sqrt{\frac{\frac{S}{I}} {\ln\left(1+\frac{S}{I}\right)}}$, which decreases with the average power of the interference.

Since massive MIMO can support high SE, a natural concern is whether the operating point of $P^*$ that achieves the maximal EE will cause a SE loss.
To address this concern, in the sequel we analyze the optimal data rate achieved by $P^*$ and the maximal achievable rate of ZFBF or MRT obtained by setting $P \to \infty$. By substituting \eqref{E:Optimal_P} into \eqref{E:Asymp_Rate_CH2_MRT} and \eqref{E:Asymp_Rate_CH2_ZFBF}, the optimal average data rate per BS with MRT and ZFBF can be obtained~as a unified expression, which is
\begin{align}
\bar{R}_{\rm{BS}}^*\approx BK\log_2\left(1+ \frac{S}{I+\frac{G}{P^*}}\right)=BK\log_2\left(1+ \frac{S}{I}\frac{1}{1+ \sqrt{\frac{(1-\frac{KT_{\rm{tr}}}{T})\eta K\sigma^2}{P_0v\alpha}(\frac{1}{S}+\frac{1}{I})\frac{\ln (1+\frac{S}{I})}{M}}}\right),\nonumber
\end{align}
and the maximal achievable  rate of ZFBF and MRT can be obtained as,
\begin{align}
\bar{R}_{\rm{BS},max}&\triangleq \lim_{P\rightarrow \infty} BK\log_2\left(1+\frac{S}{I+\frac{G}{P}}\right)=BK\log_2\left(1+\frac{S}{I}\right).\nonumber
\end{align}
By substituting $I=I_{nP}^{\rm{M}}+I_P^{\rm{M}}$ and $S=S^{\rm{M}}$ for MRT and $I=I_{nP}^{\rm{Z}}+I_P^{\rm{Z}}$ and $S=S^{\rm{Z}}$ for ZFBF, the gap between the optimal data rate and maximal achievable rate per BS can be derived as
\begin{align}\label{E:RateGap}
\Delta_R&\triangleq \bar{R}_{\rm{BS},\max}-\bar{R}_{\rm{BS}}^* \nonumber\\
&\approx BK\log_2\left(1+\sqrt{\frac{(1-\frac{KT_{\rm{tr}}}{T})\eta K\sigma^2}{P_0v\alpha}\left(\frac{1}{S}+\frac{1}{I}\right)\frac{\ln (1+\frac{S}{I})}{M}}\right),\\
&\approx
\begin{cases}
BK\log_2\left(1+\sqrt{\frac{(1-\frac{KT_{\rm{tr}}}{T})\eta K\sigma^2}{P_0v\alpha }\frac{\ln M-\ln I_{nP}}{I_{nP}M}}\right),\text{without PC},\\
BK\log_2\left(1+\frac{1}{M}\sqrt{\frac{(1-\frac{KT_{\rm{tr}}}{T})\eta K\sigma^2}{P_0v\alpha}\left(1+\tfrac{1}{\chi(\bar{L}_P-1)+\frac{I_{nP}}{M}} \right)\ln (1+\tfrac{1}{\chi(\bar{L}_P-1)+\frac{I_{nP}}{M}})}\right),  \text{with PC},
\end{cases}\label{E:RateGap1}
\end{align}
where $I_{nP}=I_{nP}^{\rm{M}}$ for MRT and $I_{nP}=I_{nP}^{\rm{Z}}$ for ZFBF, the first approximation comes from $\log_2(1+\gamma)\approx \log_2(\gamma)$ and is accurate for large SINR $\gamma$, the second approximation is from $S\approx M$ for large value of $M$, and both are accurate for massive MIMO systems.

From \eqref{E:RateGap1}, it can be observed that $\Delta_R$ decreases with the increase of $P_0$ and $M$. When $M \to \infty$, $\Delta_R \to 0$. This give rises to the following proposition.

\textbf{Proposition 2:} For the massive MIMO system with given numbers of antennas and users, the optimal average rate achieved by $P^*$ is close to the maximal average rate achieved by $P\rightarrow \infty$.

This implies that supporting the maximal EE of a massive MIMO system with given values of $M$ and $K$  by configuring transmit power will cause a little loss of the maximal sum rate achieved by ZFBF or MRT.

\emph{Remark 3}: By setting $L =1$ and $\gamma_{\rm tr} \to \infty$ in ZFBF, it is easy to show from \eqref{E:Asymp_ICI_noPC_ZF} that $I^{\rm{Z}}=I_P^{\rm{Z}}+I_{nP}^{\rm{Z}}=0$ and from \eqref{E:RateGap} that $\Delta_R \to \infty$. This means that for single cell massive MIMO system with perfect channel and ZFBF, the optimal rate is far from the maximal achievable rate (that is infinite). By contrast, for multi-cell massive MIMO system with channel estimation errors but without PC, either MUI or non-coherent ICI will cause $\Delta_R \to 0$ for large $M$. When PC is in presence,  $\Delta_R \to 0$, but with a faster decreasing speed with $M$ compared with the system without PC. In other words, in multi-cell massive MIMO systems, $\bar{R}_{\rm{BS}}^* \approx \bar{R}_{\rm{BS},\max}$.

%\emph{Remark 4}: When the average rate $\bar{R}_{\rm{BS}}$ is low, the transmit power is low, then the overall power dominates by the circuit power. In this case, $\mathrm{EE} \approx \frac{\bar{R}_{\rm{BS}}}{MP_0}$, which increases with $\bar{R}_{\rm{BS}}$ linearly. When $\bar{R}_{\rm{BS}}>\bar{R}_{\rm{BS}}^*$, the average EE begins to decrease with $\bar{R}_{\rm{BS}}$. According to Proposition 2, $\bar{R}_{\rm{BS}}^*$ is very close to $\bar{R}_{\rm{BS},\max}$, which implies that the average EE will reduce with $\bar{R}_{\rm{BS}}$ sharply. In other words, if the average rate of a massive MIMO system is less than $\bar{R}_{\rm{BS}}^*$, the achieved EE increases with the average rate. Otherwise, a little increase of $\bar{R}_{\rm{BS}}$ will induce a severe decrease of $\rm{EE}$.

To quantify how the optimal transmit power $P^*$ should increase with $M$ and compare with existing results  \cite{EESEuplink,WCNC14}, in the sequel we show its scaling law.
By substituting $I=I_{nP}^{\rm{M}}+I_P^{\rm{M}}$ and $S=S^{\rm{M}}$ into \eqref{E:Optimal_P} for MRT and $I=I_{nP}^{\rm{Z}}+I_P^{\rm{Z}}$ and $S=S^{\rm{Z}}$ into \eqref{E:Optimal_P} for ZFBF, the power scaling law  can be obtained after some regular manipulations.

\textbf{Proposition 3:} For both MRT and ZFBF, the optimal transmit power scales with $M$ as follows
\begin{equation}\label{E:Optimal_P_Scaling}
\lim_{M\rightarrow \infty}P^*\approx
\begin{cases}
\sqrt{\frac{P_0K\sigma^2(\frac{1}{\rho \gamma_{\rm{tr}}\alpha}+\bar{L}_P)}{(1-\frac{KT_{\rm{tr}}}{T})\eta \alpha I_{nP}}\frac{M}{\ln M}}\ \sim\ \mathcal{O}(\sqrt{\frac{M}{\ln M}}),\ \ \ \ \ \ \ \ \ \ \  \text{without PC},\\
\sqrt{\frac{P_0K\sigma^2(\frac{1}{\rho \gamma_{\rm{tr}}\alpha}+\bar{L}_P)}{(1-\frac{KT_{\rm{tr}}}{T})\eta \alpha}\frac{\frac{1}{\chi(\bar{L}_P-1)}-\frac{1}{1+\chi(\bar{L}_P-1)}} {\ln (1+\frac{1}{\chi(\bar{L}_P-1)})}}\ \sim\ \mathcal{O}(1),\ \  \text{with PC},
\end{cases}
\end{equation}
where $I_{nP}=I_{nP}^{\rm{M}}$ for MRT and $I_{nP}=I_{nP}^{\rm{Z}}$ for ZFBF.

\emph{Remark 4:} For the single-cell massive MIMO system with ZFBF and perfect channel information at the BS, it was shown in \cite{WCNC14} that the optimal transmit power grows proportional to $\frac{M}{\ln M}$. By setting $L=1$ in the expression in \eqref{E:Optimal_P_Scaling} without PC, the scaling law becomes $\sqrt{\frac{M}{\ln M}}$.
This indicates  that for the single-cell massive MIMO system  with ZFBF, the imperfect CSI will change the scaling law. Moreover, Proposition 3 indicates that the optimal transmit power increases with $\sqrt{\frac{M}{\ln M}}$ instead of $\frac{M}{\ln M}$ once the system suffers from the interference other than PC, and approaches to a constant independent of $M$ once the system suffers from PC. Noting that such a power scaling law is very different from that in \cite{EESEuplink} because we consider the overall power consumption at the BS but only transmit power is taken into account in \cite{EESEuplink}.\vspace{-3mm}

\subsection{Maximal EE for MRT}
By substituting $S^{\rm{M}}$, $I^{\rm{M}}=I_{nP}^{\rm{M}}+ I_{P}^{\rm{M}}$ and \eqref{E:Optimal_P} into the objective function in \eqref{E:OptimizeEE}, we can derive the approximate maximal EE of the system with MRT~as
\begin{align}\label{E:Optimal_EE_MRT}
\mathrm{EE}^{\rm{M}*}&\approx \frac{(1-\frac{KT_{\rm{tr}}}{T})BK}{P_0}\cdot \frac{\log_2\left(1+\frac{M+\frac{1}{\gamma_{\rm{tr}}\alpha} +(\bar{L}_P-1)\rho} {\chi (\bar{L}_P-1)(M-\rho) +(K\bar{L}-1)(\rho \bar{L}_P+\frac{1}{\gamma_{\rm{tr}}\alpha}) +\frac{c}{f_1(M)}}\right)}{M+cf_1(M)},
\end{align}
where $c=\sqrt{\frac{(1-\frac{KT_{\rm{tr}}}{T})\eta K\sigma^2(\frac{1}{\rho \gamma_{\rm{tr}}\alpha}+\bar{L}_P)}{ P_0}}$ is a constant, and $f_1(M)=\sqrt{\frac{\frac{M}{I^{\rm{M}}}-\frac{M} {S^{\rm{M}}+I^{\rm{M}}}}{\ln \left(1+\frac{S^{\rm{M}}} {I^{\rm{M}}}\right)}}$.

In what follows, we analyze the scaling law of maximal EE with $M$ for the massive MIMO system without and with PC, respectively.
\subsubsection{Without PC}
For the system without PC, $\bar{L}_P=1$, and $I_{P}^{\rm{M}}=0$, then $\lim_{M\rightarrow \infty}f_1(M)=\sqrt{\frac{M}{I_{nP}^{\rm{M}}\ln M}}$ and the maximal EE scales with $M$~as
\begin{equation}\label{E:MaximalEE_noPC_MRT_scaling}
\lim_{M\rightarrow \infty}\mathrm{EE}^{\rm{M}*}\approx \frac{(1-\frac{KT_{\rm{tr}}}{T})BK}{P_0} \frac{\log_2(M)-\log_2(I_{nP})}{M}\approx \frac{(1-\frac{KT_{\rm{tr}}}{T})BK}{P_0}\cdot \frac{\log_2(M)}{M},
\end{equation}
which indicates that the maximal EE decreases with $M$ asymptotically.

\subsubsection{With PC}
For the system with PC, $I_P^{\rm{M}}>0$, the maximal EE scales with $M$~as
\begin{align}\label{E:MaximalEE_PC_MRT_scaling}
\lim_{M\rightarrow \infty}\mathrm{EE}^{\rm{M}*}&\approx \frac{(1-\frac{KT_{\rm{tr}}}{T})BK}{P_0}\cdot \frac{\log_2(1+\frac{1}{\chi (\bar{L}_P-1)})-\log_2(1+\frac{I_{nP}}{\chi(\bar{L}_P-1)M})}{M} \nonumber\\ &\approx \frac{(1-\frac{KT_{\rm{tr}}}{T})BK\log_2(1+\frac{1}{\chi (\bar{L}_P-1)})}{P_0}\cdot \frac{1}{M},
\end{align}
which indicates that the maximal EE also decreases with $M$ asymptotically but with a faster speed than the system without PC.

\vspace{-3mm}
\subsection{Maximal EE for ZFBF}
By substituting $S^{\rm{Z}}$, $I^{\rm{Z}}= I_{nP}^{\rm{Z}}+I_P^{\rm{Z}}$ and \eqref{E:Optimal_P} into the objective function in \eqref{E:OptimizeEE}, the approximate maximal EE of the system with ZFBF can be derived as
\begin{align}\label{E:Optimal_EE_ZF}
\mathrm{EE}^{\rm{Z}*}\approx \frac{(1-\frac{KT_{\rm{tr}}}{T})BK}{P_0}\frac{\log_2\left(1+ \frac{M+\frac{1}{\gamma_{\rm{tr}}\alpha}+(\bar{L}_P-1-K)\rho}{\chi (\bar{L}_P-1)(M-2\rho K)+(K\bar{L}-1)(\rho\bar{L}_P+\frac{1}{\gamma_{\rm{tr}}\alpha})-\rho(K-1) +\frac{c}{f_2(M)}}\right)}{M+cf_2(M)},
\end{align}
where $f_2(M)= \sqrt{\frac{\frac{M}{I^{\rm{Z}}} -\frac{M}{S^{\rm{Z}}+I^{\rm{Z}}}}{\ln \left(1+\frac{S^{\rm{Z}}}{I^{\rm{Z}}}\right)}}$.

With (\ref{E:Optimal_EE_ZF}) and using similar method as analyzing MRT, we can obtain the scaling law of the maximal EE for the systems with ZFBF. For conciseness, the derivation is omitted, where the results show that the systems with MRT and ZFBF can achieve exactly the same scaling law.

\subsection{Summary of the Results}

The scaling laws of the optimal transmit power and maximal EE of the massive MIMO systems with respect to $M$ are summarized in Table \ref{T:Scalinglaw}.

\begin{table}[!htb]
\caption{Scaling Law of Optimal Transmit Power and Maximal EE with Number of Antennas for MRT and ZFBF }\label{T:Scalinglaw}\centering
\begin{tabular}{|l|l|l|}
\hline
 & $P^{*}$ for MRT and ZFBF, $M\rightarrow \infty$, given $K$& $\mathrm{EE}^{*}$ for MRT and ZFBF, $M\rightarrow \infty$, given $K$ \\
\hline
Without PC& $\sqrt{\dfrac{P_0K\sigma^2(\frac{1}{\rho \gamma_{\rm{tr}}\alpha}\!+\!\bar{L}_P)}{(1-\frac{KT_{\rm{tr}}}{T})\eta \alpha I_{nP}}}\sqrt{\dfrac{M}{\ln M}}\sim \mathcal{O} \left(\sqrt{\dfrac{M}{\ln M}}\right)$& $\dfrac{(1-\frac{KT_{\rm{tr}}}{T})BK}{P_0}\dfrac{\log_2M}{M} \sim \mathcal{O}\left(\dfrac{\log_2 M}{M}\right)$\\
\hline
With PC &$\sqrt{\dfrac{P_0K\sigma^2(\frac{1}{\rho \gamma_{\rm{tr}}\alpha}\!+\!\bar{L}_P)}{(1-\frac{KT_{\rm{tr}}}{T})\eta \alpha}\dfrac{\frac{1}{\chi(\bar{L}_P-1)}\!-\!\frac{1}{1+\chi(\bar{L}_P-1)}} {\ln (1+\frac{1}{\chi(\bar{L}_P-1)})}}\sim \mathcal{O}(1)$& $\dfrac{(1\!-\!\frac{KT_{\rm{tr}}}{T})BK\log_2(1\!+\!\frac{1}{\chi(\bar{L}_P-1)})} {P_0}\dfrac{1}{M}\sim \mathcal{O}\left(\dfrac{1}{M}\right)$ \\
\hline
\end{tabular}
\end{table}

The results can be explained as follows. When the number of users $K$ is given, for the system without PC, the optimal sum rate increases with $\log_2 M$ owing to the increased array gain, while for the system with  PC, the optimal rate  approaches to a constant independent of $M$ owing to the coherent ICI. On the other hand, the overall circuit power consumption  increases with $M$. As a consequent, the maximal EE of massive MIMO system with MRT or ZFBF finally decreases with $M$. Due to the training overhead that depending on $K$, the maximal EE does not simply increase with the multiplexing gain of $K$. In fact, as have been analyzed in \cite{WCNC14,EELSAS_Marzetta13}, there exists an optimal number of users for fixed values of $M$ and $P$ and a fixed value of $M/K$.

Based on the table and previous analysis, we can answer the proposed questions as follows.

(1) Given the number of users, the maximal EE decreases with the increase of $M$, no matter if the massive MIMO system is with or without PC.

(2) Given the number of users, in order to maximize the EE, the transmit power should be configured to increase with $M$  according to the laws in Table \ref{T:Scalinglaw}. Since for massive MIMO system the maximal achievable rate increases with $M$ but the maximal EE decreases with $M$, meanwhile $\bar{R}_{\rm{BS}}^* \approx \bar{R}_{\rm{BS},\max}$ as addressed in \emph{Remark 3}, an energy efficient practical system should configure $M$ as the minimum number of antennas that can support a required average rate, which depends on the traffic load in a specific time and location.
%set a minimal rate requirement for the non-real-time services such as file transfer, and $M$ should be configured as the minimum number of antennas that can support the required minimal average rate.
%If a system has a minimal rate requirement (say, when the system serves the users with non-real-time services such as file transfer),
%the maximal EE decreasing with $M$ implies that the EE-maximal massive MIMO should be configured with the minimum number of antennas that can support the minimal average rate.

(3) In single-cell scenario, the channel estimation errors change the scaling law of $P^{*}$ for massive MIMO system with ZFBF. In multi-cell scenario, PC changes the scaling laws of $P^{*}$ and $\mathrm{EE}^{*}$ for both massive MIMO systems with MRT and ZFBF. Channel correlation affects the power scaling law and the impact for the systems with MRT and with ZFBF differs, but does not affect the EE scaling law.

%The circuit power consumption per antenna should decrease with $M$ as $\frac{\log_2 M}{M}$ for the system without PC and $\frac{1}{M}$ for the system with PC. The system with PC should work at the maximal EE point which will achieve approximate maximum SE due to the coherent ICI, but the system without PC should work at the maximal SE point. This is because when circuit power decreases with $\frac{\log_2M}{M}$, the optimal transmit power is a constant independent with $M$ and plays an unimportant role in the total power consumption. Therefore, the system transmitting at the maximum transmit power will achieve approximate maximal EE.

\section{Numerical and Simulation Results}
In this section, we validate previous analytical results via simulations.

Consider a massive MIMO system consisting of seven macro cells each with radius 250~m, where one central cell  is
surrounded by six cells. In each cell 10 users are located randomly with a minimum distance of 35~m from the BS.
The path loss model is set as 35.3+37.6$\log_{10}d$~dB, where $d$ (in~m) is the distance between the user and BS.
Both the uplink and downlink noise variances are  -174~dBm/Hz\cite{TR36.814}, and the system bandwidth is 20 MHz. The length of a coherence block is
$T=T_cB_c =$ 1500 channel uses, where $T_c=$ 3.8~ms is the coherence time for 60~km/h moving speed and 2~GHz
carrier and $B_c =$ 400~kHz is the coherence bandwidth for urban macro cell \cite{TR36.814}. In the uplink
training phase, the transmit power is $P_{\rm{tr}} =$ 200 mW, and $T_{\rm{tr}}=1$ channel use. MMSE channel estimator is used. Since in the
literature there are no specific power consumption parameters for a massive MIMO BS, we evaluate the EE of
massive MIMO systems with parameters in the year 2012 and the predicated values in the year 2020 provided by
GreenTouch consortium, which are respectively $P_c=1.42$~W and $P_{sp}=3.1$~mW (2012) and $P_c=0.2$~W and
$P_{sp}=$ 0.4~mW (2020)\cite{GreenTouch_Power}, and $\eta =$ 2.51 obtained from the PA efficiency $\eta_{\rm
PA}=50$\% \cite{GreenTouch_Power} and loss factors $\sigma_{\rm{DC}}=6$\%, $\sigma_{\rm{cool}}=9$\% and
$\sigma_{\rm{MS}}=7$\% for macro BS \cite{E3F_EARTH} are the same for both years of 2012 and 2020. The
performance of the central cell is evaluated. Unless otherwise specified, these simulation setups  will be used
throughout the simulations.

\begin{figure}
\centering
\includegraphics[width=0.55\textwidth]{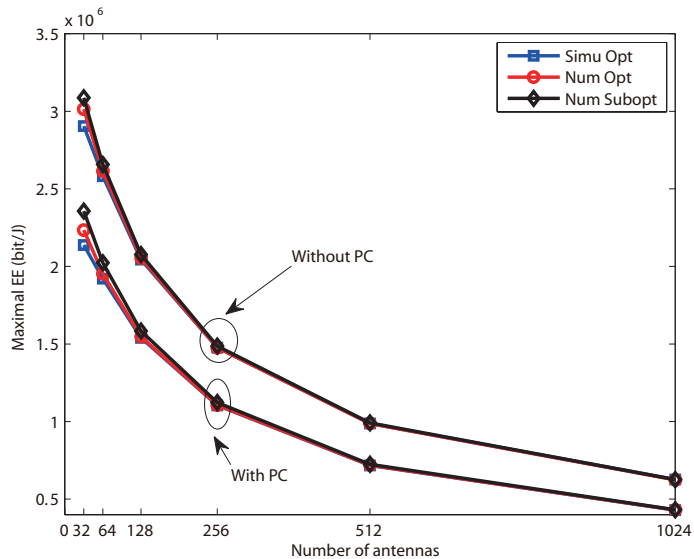}
\caption{Evaluation of the approximations, $P_c=1.41$~W and $P_{sp}=3.1$~mW (2012), with MRT.} \label{F:MaximalEE_vs_Nt_validate}
\end{figure}
\subsection{Accuracy of the Approximations}
To show the impact of the approximations on the analysis, we compare the maximal EEs respectively obtained from the following approaches in Fig.~\ref{F:MaximalEE_vs_Nt_validate}, where the channel model in \eqref{E:Channel_Model} with $\alpha=\frac{10^{-3.53}}{250^{3.76}}$, $\rho=2$ and $\chi=0.1$ is used.
\begin{itemize}
    \item[a)]  The optimal solution with legend ``Simu Opt'' is obtained through exhaustive searching $P^*$ from an EE maximization problem without any approximations, where the average data rate per user is obtained from simulation by averaging over small scale channels.
  \item[b)]  The numerical results with legend ``Num Opt'' are found respectively from optimization problem \eqref{E:OptimizeEE} for the system using MRT and ZFBF by bisection searching, where the asymptotic data rates are used.
  \item[c)]  The approximated solutions with legend ``Num SubOpt'' are computed with  the approximated maximal EEs in \eqref{E:Optimal_EE_MRT} for MRT and \eqref{E:Optimal_EE_ZF} for ZFBF.
\end{itemize}

The results show that the approximations are accurate for massive MIMO systems with MRT. The results for the system with ZFBF  are similar and are not shown to make the figure  clear.

\vspace{3mm}In order to demonstrate that the analytical results are also valid for more realistic channels, in the rest of the subsections we simulate the performance of massive MIMO systems by averaging over large-scale and small-scale channels, where a three-dimensional (3D) MIMO system with uniform rectangular array in urban macro (UMa) scenario  (referred as 3D UMa) is employed \cite{TR36.873}, considering that such a large number of antennas may not be arranged as a linear array. Note that the lognormal distributed shadowing with 4 dB deviation is considered in the 3D UMa channel, and the users are randomly placed in each cell. In the simulation, PC will be automatically generated in the massive MIMO systems unless otherwise specified. The antenna spacing at the BS is half of the wavelength for both horizontal and vertical directions. The main 3D MIMO channel parameters are listed in Table \ref{T:3Dchannel}.
%
%\subsection{Validation of the Analytical Results}
%To show the impact of the employed channel model on the asymptotic analysis results, we compare the average data rate per BS respectively obtained from the following approaches.
%\begin{itemize}
%  \item[a)]  \emph{Sim-3D Uma:} The average data rate is obtained via simulations by averaging over large-scale and small-scale channels, where a three-dimensional (3D) MIMO system with uniform rectangular array in urban macro (UMa) scenario with NLOS from an outdoor BS to an indoor user (referred as 3D UMa) is employed \cite{TR36.873}. Note that the lognormal distributed shadowing with 4 dB deviation is considered in 3D UMa. The antenna spacing at the BS is half of the wavelength for both horizontal and vertical directions. The main 3D MIMO channel parameters are listed in Table \ref{T:3Dchannel}.
%  \item[b)]  \emph{Sim-Co CH:} The average data rate is obtained via simulations by averaging over small-scale channels, where the correlated channel model in \eqref{E:Channel_Model} is employed with $G_c=10^{-3.53}250^{-3.76}$, $\alpha=0.1$ and $\beta=2$.
%  \item[c)]  \emph{Num-Co CH:} The asymptotic data rate is computed with  \eqref{E:Asymp_Rate_CH2_MRT} for MRT and \eqref{E:Asymp_Rate_CH2_ZFBF} for ZFBF, respectively.
%\end{itemize}
%
\begin{table}[!htb]
\vspace{-5mm}
\caption{Main Parameters of 3D MIMO Channels}\label{T:3Dchannel}\centering
\begin{tabular}{c|l|c|l}
\hline
\hline
    Mean of azimuth clusters & $U(-60^\circ,60^\circ)$&
    Mean of elevation clusters & $U(-45^\circ,45^\circ)$\\
\hline
Mean of log delay spread (DS) ($\log_{10}([s])$)&-6.62&
Variance of log DS  & 0.32 \\
\hline
Mean of log azimuth spread (AS) ($\log_{10}([^\circ])$)&1.25&
Variance of log AS &0.42\\
\hline
Mean of log elevation spread (ES) ($\log_{10}[^\circ]$)&$\max[-0.5,-2.1\frac{d}{1000}+0.9]$&
Variance of log ES &0.49\\
\hline
Number of clusters&12&
Number of rays per cluster&20\\
\hline
\hline
\end{tabular}
\end{table}
%
%\begin{figure}
%\centering
%\includegraphics[width=0.55\textwidth]{AverageRate_vs_Nt}
%\caption{Average data rate per BS versus $M$, where $P=$ 40~W. The same set of pilot sequences is reused in seven cells.} \label{F:AverageRate_vs_Nt}
%\end{figure}
%
%In Fig.~\ref{F:AverageRate_vs_Nt}, we show the rate versus the number of antennas. It is shown that when the channel model given in \eqref{E:Channel_Model} is considered, the numerical results ``Num-CH2'' overlap with the simulation results ``Sim-CH2'', which indicates that the asymptotic analysis in deriving the average data rate is accurate even for not so large number of antennas and users, e.g. $M=64$ and $K=10$. The data rate in ``Sim-CH1'' is higher than that in ``Sim-CH2'', because the former considers random located users within each cell while the later considers the users located at cell-edge..
%
%To demonstrate that our conclusions in the analysis with 'CH2' are also valid for more realistic channel model, we simulate the performance of 'CH1' using the 3D UMa channel in the sequel.

\vspace{-3mm}\subsection{EE Comparison Between Massive MIMO and LTE Systems}
To validate the analytical results and compare the EE of massive MIMO with traditional MIMO systems (say, LTE systems), we show the EE-Rate
curves under various settings.
The average data rate per BS with MRT or ZFBF and corresponding EE of the system are obtained from simulations
under 3D UMa channel, where the transmit power $P$ increases from 0 W without limit and $P*$ is found by
exhaustive searching. To show the feasible region of the rate achievable by a macro BS with a maximal transmit
power of 40 W (the maximal transmit power for a LTE macro BS), the results with $P > 40$ W are plotted with
dotted curves.

\begin{figure}
\centering
\includegraphics[width=0.6\textwidth]{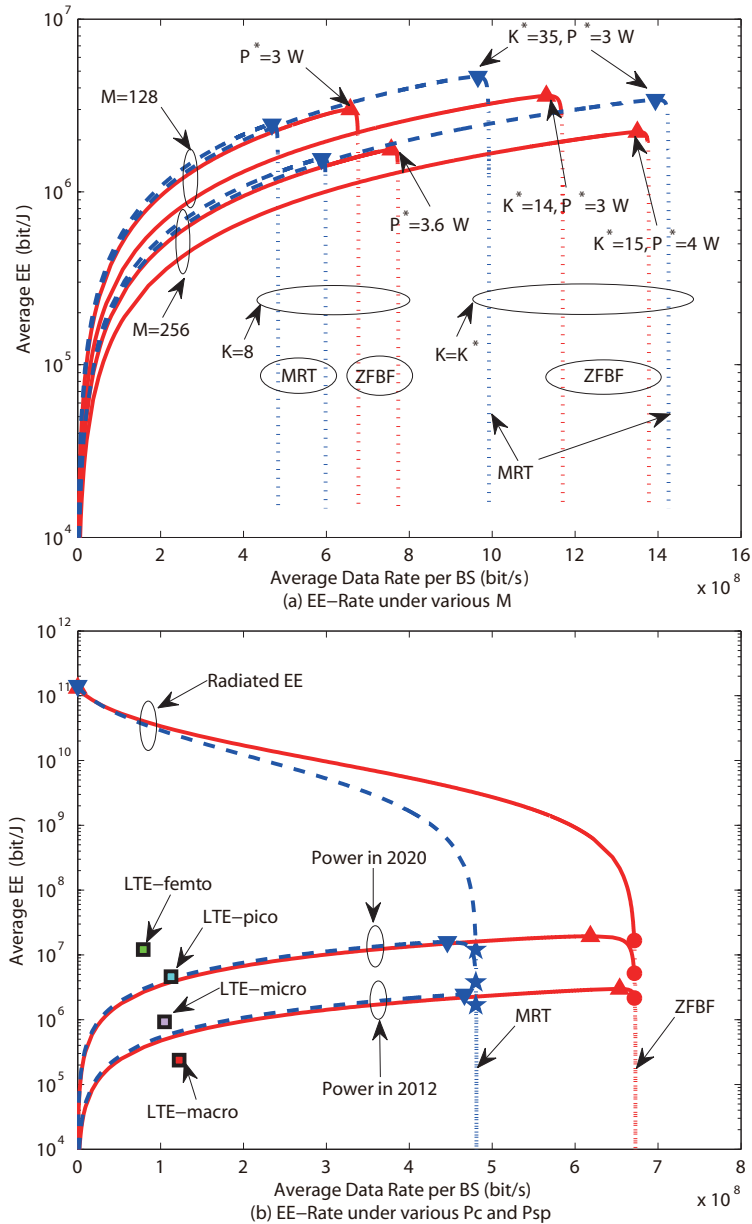}
\caption{EE-Rate relationship under various $M$ and $P_c$, $P_{sp}$. The dashed and solid lines represent the results for MRT and ZFBF, and the maximal EEs for MRT and ZFBF are marked by lower and upper triangle, respectively. In Fig.~\ref{F:EE_vs_Rate}~(a), $P_c=$ 1.42~W and $P_{sp}=$ 3.1~mW (for the year of 2012). In Fig.~\ref{F:EE_vs_Rate}~(b), $M = 128$ for the massive MIMO BS, the EEs
achieved by $P$ = 40~W are marked by stars for MRT and bullets for ZFBF, and the achieved EE-Rate point by four types of LTE BSs are marked by squares. In both Figs.~\ref{F:EE_vs_Rate}~(a) and (b), $L=7$, 3D UMa channel model is employed for the massive MIMO system and i.i.d. channel model is employed for LTE systems. }\label{F:EE_vs_Rate}
\end{figure}

In Fig.~\ref{F:EE_vs_Rate}~(a), we show the impact of the number of transmit antennas $M$  as well as  the impact of optimizing the transmit power and optimizing the number of users, where the power
consumption parameters provided by GreenTouch consortium for the year 2012 are considered (the results with the
parameters for 2020 are similar and hence are not shown).

We first fix the number of users, considering the application scenario with low traffic load where user scheduling is no necessary. To be comparable with at least one kind of LTE BS, say macro BS with $M=8$, we set $K=8$. Due to the MUI, non-coherent ICI and coherent ICI, the
data rates of the massive MIMO systems are limited to finite values when $P \to \infty$ for both MRT and ZFBF, as
addressed in \emph{Remark 2}. The maximal EEs are achieved approximately at the maximal achievable rates of ZFBF and MRT,  which agrees with  \emph{Remark 3}. Before the sharp reduction, the EEs
achieved by MRT and ZFBF are very close, which can be interpreted by \emph{Remark 1} and their similar level of
circuit powers (with only difference in $P_{sp}$). The maximal EE achieved by ZFBF exceeds that by MRT due to the higher optimal rate, but their maximal EEs will finally become identical (not shown in the figure) again due to the fact in \emph{Remark 1}. The maximal EEs decrease with $M$, and the optimal transmit powers are much less than 40~W. For example, for massive MIMO system with ZFBF, when $M$ = 256,  $P^*$ = 3.6 W. The achieved EE is sensitive to the transmit power, especially when it is configured beyond the optimal value.

In practice, there may exist large number of users in a macro cell,  e.g., in a hot-spot area during busy hours in
a day. In such a scenario, we can further optimize $K$ together with $P$ to maximize EE, which is achievable by
using an optimal user scheduler. In Fig.~\ref{F:EE_vs_Rate}~(a) we also give the EE-Rate relationship of the
system with the optimal number of users $K^*$ (see the group of curves marked by ``$K=K^*$''), where $K^*$ is obtained by jointly optimizing $(K,P)$ to maximize
the average EE when the number of antennas $M$ is given. Specifically, the average EE is obtained by simulation
under 3D UMa channel, and $(K^*,P^*)$ is found by exhaustive searching. It can be observed that $K^*$ for the
system with MRT is larger than that with ZFBF, and both exceed $K=8$.
%In the low rate region, the system with MRT
%for a given $M$ with $K^*$ achieves almost the same EE as that with $K=8$, while the system with ZFBF  for a
%given $M$ with $K^*$ achieves lower EE than that with $K=8$ due to the increased signal processing power $P_{{\rm
%BB}_2}$. In the high rate region,
The system for a given $M$ with $K^*$ achieves higher EE than the system with $K=8$ as a result of higher
rate, both for ZFBF and MRT, while optimizing $K$ brings higher EE gain for MRT than for ZFBF. When $M=256$, the
system with MRT and $K^*$ (which equals to 35) can achieve both higher maximal EE and higher achievable rate than the system with
ZFBF and $K^*$ (which equals to 15). The maximal EE still decreases  with $M$, which indicates that optimizing the number of users
will not change the scaling law of the maximal EE with $M$.

\begin{table}[!htb]
\caption{System Parameters of Four Types LTE Systems \cite{E3F_EARTH} (``PL": path loss model, $D_c$: cell radius)}\label{T:LTEparamters}\centering
\begin{tabular}{c|l}
\hline
    BS type&System and Power Consumption Parameters \\
\hline
    Macro&$M\!=\!8,K\!=\!8,D_c\!=\!250\text{m},\text{PL}\!=\!\frac{10^{-3.53}}{d^{3.76}}, P_{\text{max}}\!=\!40\text{W},P_c\!=\!31.7\text{W},P_{sp}\!=\!7.8\text{mW},\eta\!=\!3.24,\sigma_{\text{feed}}\!=\!-3\text{dB}$\\
\hline
    Micro&$M\!=\!4,K\!=\!4,D_c\!=\!100\text{m},\text{PL}\!=\!\frac{10^{-3.53}}{d^{3.76}}, P_{\text{max}}\!=\!6.3\text{W},P_c\!=\!21.4\text{W},P_{sp}\!=\!23.5\text{mW},\eta\!=\!4.04,\sigma_{\text{feed}}\!=\!0$\\
\hline
    Pico&$M\!=\!4,K\!=\!4,D_c\!=\!50\text{m},\text{PL}\!=\!\frac{10^{-3.06}}{d^{3.67}}, P_{\text{max}}\!=\!0.13\text{W},P_c\!=\!2.6\text{W},P_{sp}\!=\!2.7\text{mW},\eta\!=\!13.72,\sigma_{\text{feed}}\!=\!0$\\
\hline
    Femto &$M\!=\!2,K\!=\!2,D_c\!=\!30\text{m},\text{PL}\!=\!\frac{10^{-3.06}}{d^{3.67}}, P_{\text{max}}\!=\!0.05\text{W},P_c\!=\!1.9\text{W},P_{sp}\!=\!7.2\text{mW},\eta\!=\!21.11,\sigma_{\text{feed}}\!=\!0$\\
\hline
\end{tabular}
\end{table}

In Fig.~\ref{F:EE_vs_Rate}~(b), we show the impact of $P_c$ and $P_{sp}$, where the circuit power consumption
parameters in 2020  and 2012  are considered, and $K$ is fixed as eight. As expected, the maximal EE reduces with the increase of the
circuit power consumption.
%In low data rate region, the maximal EE for MRT is slightly higher than ZFBF because
%of lower computational complexity. In high data rate region,
ZFBF achieves a higher maximal EE as a result of
higher achievable rate. Along with the increase of circuit power consumption, the rate corresponding to the the EE-maximizing optimal transmit
power  moves closer to the point with infinite transmit power for both MRT and ZFBF, which agrees with \emph{Proposition 2}. For comparison, the EE-Rate points achieved by four types of LTE systems using
ZFBF transmitting at their corresponding values of $P_{\text{max}}$ are provided, where the parameters given in
Table \ref{T:LTEparamters} are for the year of 2012, and  the radiated EE \cite{EESEuplink} is also shown, where $P_c =$ 0
and $P_{sp}=$ 0 and $\eta =$ 1. As suggested in \cite{EELSAS_Marzetta13} as well as indicated by the EE scaling laws (i.e., EE is inversely proportional to $P_0$), the EE improvement largely depends on the value of the circuit powers. We can see that the maximal EE of the massive MIMO system with ZFBF and predicted power
consumption parameters in 2020 is higher than all LTE systems (e.g., 82 times higher the LTE macro system), and that with the parameters in 2012 is only higher than the LTE-micro and macro systems (e.g., 13 times
higher than the LTE macro system).

We can also obtain a target maximal EE by changing the equivalent circuit power consumption per-antenna $P_0$, from which we can see how the value of $P_0$ should be to achieve an expected fold of EE improvement over traditional systems. For example, for the LTE macro system in 2010, the EE is $\rm{EE}_{LTE}=9.13\times 10^4$ bit/Joule, where the
cell spectral efficiency is 6.16~bps/Hz/cell \cite{TR36.814} and the power consumption is 1350~W
\cite{E3F_EARTH}. If a massive MIMO system with 256 or 512 antennas serving 10 users without PC is expected to achieve
$1000 \cdot \rm{EE}_{LTE}$, the required value of $P_0$ should be 28~mW (if $M =256$) or 19~mW (if $M=512$) for MRT and 37~mW  (if $M =256$) or
23~mW  (if $M =512$) for ZFBF, which are much smaller than the predicted circuit power per-antenna $204$~mW for MRT and 244~mW for ZFBF
in 2020 by GreenTouch \cite{GreenTouch_Power}.

\subsection{Impact of Number of Antennas on Optimal Transmit Power}

\begin{figure}
\centering
\includegraphics[width=1.05\textwidth]{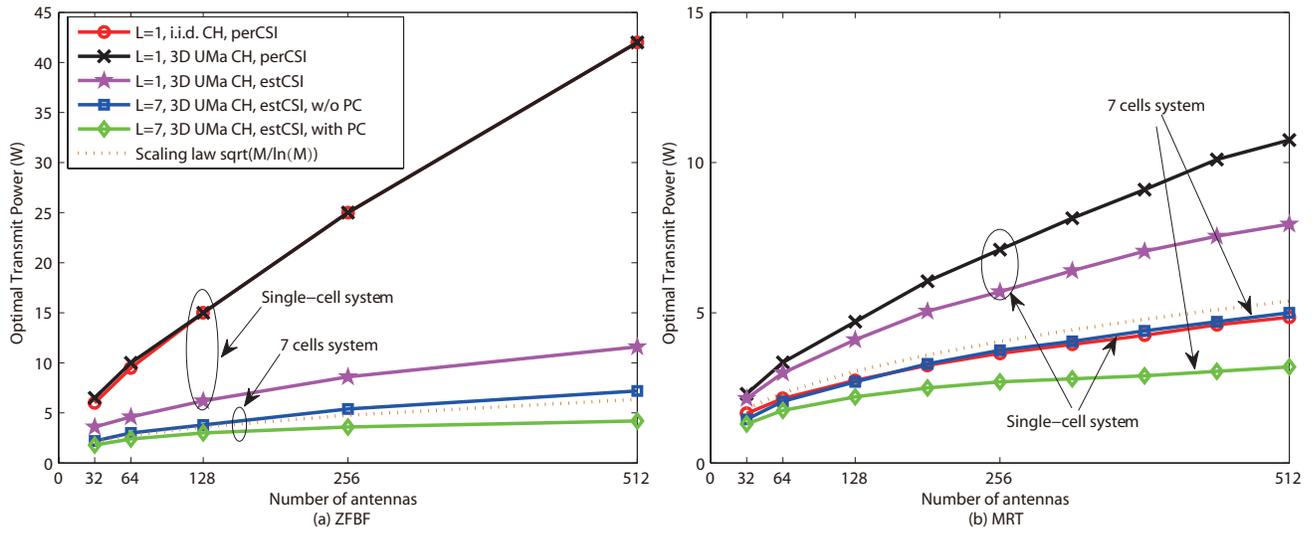}
\caption{Optimal transmit powers of massive MIMO systems with ZFBF and MRT versus $M$, the legends ``perCSI'' and ``estCSI'' respectively denote the results with perfect channel information and estimated channels. } \label{F:OptimalPt_vs_Nt}
\end{figure}

The optimal transmit power versus $M$ are given in Fig.~\ref{F:OptimalPt_vs_Nt}. We simulate the optimal transmit power in five scenarios, which can observe the impacts of channel correlation, channel estimation errors, non-coherent ICI and coherent ICI, respectively. The circuit power parameters for the year of 2012 are considered, while the results of the system with the values for 2020 are similar.

The scaling laws of the system with ZFBF are given in Fig.~\ref{F:OptimalPt_vs_Nt}~(a). We can see that  the optimal transmit power increases with $M$. For the single-cell system with perfect CSI, the optimal transmit power with i.i.d. and 3D UMa channels are nearly the same for ZFBF, which indicates that spatially correlated channel has little impact on the power scaling law of the system with ZFBF. When the channel estimation errors or non-coherent ICI is considered, the optimal transmit power increases with $M$ in the law of $\sqrt{\frac{M}{\ln M}}$ and increases much slower than the system with perfect channel information, which is consistent with \emph{Remark 4}. When the coherent ICI is considered, the optimal transmit power approaches to a constant for large $M$. As expected, the optimal transmit power decreases when the system suffers more interference. The scaling laws in different scenarios are consistent with \emph{Proposition 3}.

The scaling laws of the system with MRT are given in Fig.~\ref{F:OptimalPt_vs_Nt}~(b). The optimal transmit power increases with $M$ proportional to $\sqrt{\frac{M}{\ln M}}$ for all scenarios except with PC, which approaches a constant. In contrast to the results for ZFBF, the optimal transmit power of the system with i.i.d channel is much smaller than that with 3D UMa channel, which means that the spatially correlated channel has large impact on the performance of the system with MRT. This is because the system under the 3D UMa channel suffers less interference than that under the i.i.d. channel.

Comparing Figs.~\ref{F:OptimalPt_vs_Nt}~(a) and (b), we can find that the optimal transmit power of the system with ZFBF is much higher than that with MRT.

\subsection{Impact of Number of Antennas on Maximal EE}

\begin{figure}
\centering
\includegraphics[width=1.05\textwidth]{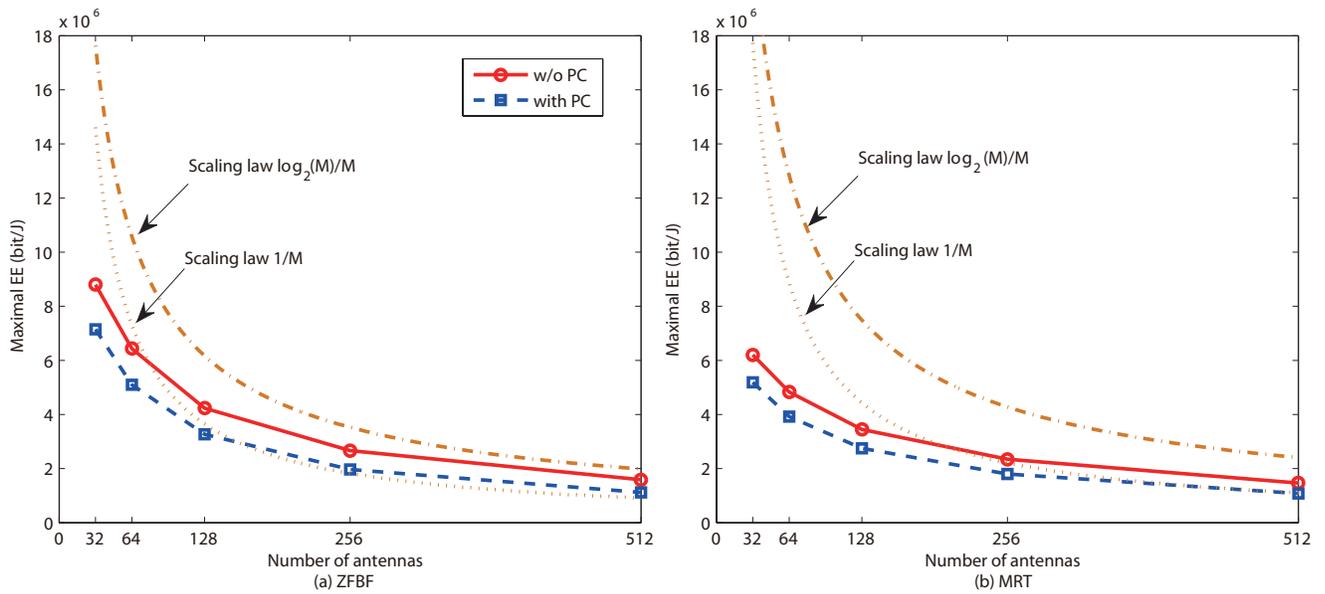}
\caption{Maximal EEs of massive MIMO systems with ZFBF and MRT versus $M$, 3D UMa channel model.} \label{F:MaximalEE_vs_Nt}
\end{figure}

The maximal EEs of the system with ZFBF and MRT versus $M$ are given in Fig.~\ref{F:MaximalEE_vs_Nt}, where the power consumption parameters in 2012 are used (the results of the system with the power consumption in 2020 are similar). It can be observed that the maximal EE decreases with $M$ and is proportional to $\frac{\log_2(M)}{M}$ for the system without PC and $\frac{1}{M}$ for the system with PC for large $M$, i.e., the maximal EE of the system with PC decreases with $M$ faster than that without PC. Moreover, the system without PC is more energy efficient than the system with PC. These results are consistent with the previous analytical results.

\section{Conclusions}
In this paper, we investigated the potential of downlink multicell massive MIMO system in improving EE by optimizing the transmit power. We derived the closed-form expressions of approximated maximal EEs of the massive MIMO systems with MRT and ZFBF under a spatially correlated channel model, and then analyzed the scaling laws of the optimal transmit power and maximal EE with the number of transmit antennas $M$ for the  system with and without pilot contamination. Analytical results showed that the maximal EE of massive MIMO systems reduces with $M$, and the decreasing speeds are different for the systems with and without pilot contamination. The optimal transmit power should be configured to increase with $M$ in order to maximize the EE of the system without pilot contamination, but be configured as a constant independent from $M$ for the system  with pilot contamination. Channel correlation has large impact on the power scaling law for massive MIMO with MRT, and has minor impact on that with ZFBF. For practical systems with given traffic loads, the most energy efficient massive MIMO system should be configured to use the minimum number of antennas that can support the required average sum rate.  Simulation results validated the analytical analysis, and showed that the conclusions are also valid under more realistic channel model.
%Moreover, with the the power
%consumption parameters predicted by GreenTouch consortium for the year of 2020, the maximal EE of the massive MIMO system will be higher than those achieved by four type LTE systems in the year of 2012.

\appendices
\section{Derivation of Asymptotic Data Rate with ZFBF}\label{P:AverageRate_ZF}
\renewcommand{\theequation}{A.\arabic{equation}}
In order to derive the asymptotic data rate with ZFBF, we first derive the asymptotic rate of the regularized zero-forcing (RZF) beamforming. The beamforming matrix of $\text{BS}_j$ with RZF~is
\setcounter{equation}{0}
\begin{equation}\label{E:W_RZF}
\mathbf{W}_{j}=\left(\mathbf{\hat{H}}_{jj}\mathbf{\hat{H}}_{jj}^H+M\varphi \mathbf{I}_M\right)^{-1}\mathbf{\hat{H}}_{jj}
\triangleq \mathbf{\Pi}_j \mathbf{\hat{H}}_{jj} =[\mathbf{w}_{j1},\ldots,\mathbf{w}_{jK}]\in \mathbb{C}^{M\times K},
\end{equation}
where $\varphi$ is the regularized parameter, $\mathbf{\Pi}_j\triangleq \left(\mathbf{\hat{H}}_{jj}\mathbf{\hat{H}}_{jj}^H+M\varphi \mathbf{I}_M\right)^{-1}$ satisfying $\mathbf{\Pi}_j^H=\mathbf{\Pi}_j$, and $\mathbf{w}_{jk}=\mathbf{\Pi}_j\mathbf{\hat{h}}_{jjk}$.
The receive SINR of $\text{UE}_{jk}$ with RZF is given~by
\begin{equation}\label{E:SINR_RZF}
\gamma_{jk}^{\rm{R}}=\frac{\bar{\lambda}_{jk} \|\frac{1}{M}\mathbf{h}_{jjk}^H \mathbf{\Pi}_{j}\mathbf{\hat{h}}_{jjk}\|^2} {\sum\limits_{l\in \mathcal{S}_j\backslash j}\bar{\lambda}_{lk}\|\frac{1}{M}\mathbf{h}_{ljk}^H \mathbf{\Pi}_{l}\mathbf{\hat{h}}_{llk}\|^2 +\sum\limits_{(l\in \mathcal{S}_j,m\neq k)\cup (l\notin \mathcal{S}_j,m)}\bar{\lambda}_{lm}\|\frac{1}{M}\mathbf{h}_{ljk}^H \mathbf{\Pi}_{l}\mathbf{\hat{h}}_{llm}\|^2 +\frac{K\sigma^2}{MP}},
\end{equation}
where $\bar{\lambda}_{jk}=M\lambda_{jk}$.

According to Theorem 3 in \cite{LargeSystem}, the asymptotic SINR of $\text{UE}_{jk}$ with RZF, $\bar{\gamma}_{jk}^{\rm{R}}$, converges to the asymptotic SINR with ZFBF, $\bar{\gamma}_{jk}^{\rm{Z}}$, when the regularized parameter approaches zero, i.e., $\bar{\gamma}_{jk}^{\rm{Z}}=\lim_{\varphi\rightarrow0}\bar{\gamma}_{jk}^{\rm{R}}$.
In the sequel, we will first derive the asymptotic SINR with RZF, and then obtain the asymptotic SINR with ZFBF by letting $\varphi\rightarrow 0$.

\subsubsection{Normalizing Parameter} According to the proof of Theorem 2 and Theorem 3 in \cite{LargeSystem}, the normalizing parameter $\bar{\lambda}_{jk}$ of ZFBF can be obtained as
\begin{equation}\label{E:Asymp_Para_ZF}
\bar{\lambda}_{jk}\xrightarrow[M,K\rightarrow \infty]{a.s.} \frac{M^2(1+\delta_j(\varphi))^2}{\delta_j^{'}(\varphi)} \xrightarrow[\varphi\rightarrow 0]{}M^2\delta^0,
\end{equation}
where the symbol $a.s.$ stands for almost surely convergence, $\delta_{j}(\varphi)=\frac{1}{M}\mathrm{tr}(\mathbf{\Phi}_{jj}\mathbf{T}_j(\varphi))$,  $\mathbf{T}_j(\varphi)=(\frac{K}{M}\frac{\mathbf{\Phi}_{jj}}{1+\delta_{j}(\varphi)} +\varphi \mathbf{I}_M)^{-1}$, $\delta_{j}^{'}(\varphi)=\frac{1}{M}\mathrm{tr}(\mathbf{\Phi}_{jj}\mathbf{T}_j^{'}(\varphi))$, $\mathbf{T}_j^{'}(\varphi)=\mathbf{T}_{j}(\varphi) \left(\mathbf{I}+\frac{K}{M} \frac{\mathbf{\Phi}_{jj}\delta_{j}^{'}(\varphi)}{(1+\delta_{j}(\varphi))^2}\right)\mathbf{T}_{j}(\varphi)$, $\mathbf{\Phi}_{jj}=v\alpha\rho\tilde{\mathbf{R}}\tilde{\mathbf{R}}^H$, and $\delta^0=v\alpha\frac{M-\rho K}{M}$.

\subsubsection{Signal Power} According to the proof of Theorem 6 in \cite{HowManyAntennas13}, the power of desired signal can be derived as
\begin{equation}\label{E:Inst_Sig_RZF}
\left|\tfrac{1}{M}\mathbf{h}_{jjk}^H\mathbf{\Pi}_{j}\mathbf{\hat{h}}_{jjk}\right|^2 =\frac{|\frac{1}{M}\mathbf{h}_{jjk}^H\mathbf{\Pi}_{jk}\mathbf{\hat{h}}_{jjk}|^2} {(1+\mathbf{\hat{h}}_{jjk}^H\mathbf{\Pi}_{jk}\mathbf{\hat{h}}_{jjk})^2},
\end{equation}
where the term $\mathbf{\hat{h}}_{jjk}^H\mathbf{\Pi}_{jk}\mathbf{\hat{h}}_{jjk}$ in the denominator asymptotically approaches to
\begin{equation}\label{E:Asymp_3}
\mathbf{\hat{h}}_{jjk}^H\mathbf{\Pi}_{jk}\mathbf{\hat{h}}_{jjk}\xrightarrow[M,K\rightarrow \infty]{a.s.}\delta_{j}(\varphi).
\end{equation}

With the expression of $\mathbf{\hat{h}}_{jjk}$ given in \eqref{E:MMSE_CHest}, the numerator of \eqref{E:Inst_Sig_RZF} becomes
\begin{align}
&|\tfrac{1}{M}\mathbf{h}_{jjk}^H\mathbf{\Pi}_{jk}\mathbf{\hat{h}}_{jjk}|^2\nonumber\\
&=
|\tfrac{1}{M}\mathbf{h}_{jjk}^H\mathbf{\Pi}_{jk}\mathbf{R}_{jj}\mathbf{Q}\mathbf{h}_{jjk}|^2 \\
&+\tfrac{1}{M^2}\mathbf{h}_{jjk}^H\mathbf{\Pi}_{jk}\mathbf{R}_{jj}\mathbf{Q} \left(\sum\nolimits_{l\in \mathcal{S}_j\backslash j}\mathbf{h}_{jlk}+\tfrac{1}{\sqrt{\rho_{\rm{tr}}}}\mathbf{n}_{jk}\right) \left(\sum\nolimits_{l\in \mathcal{S}_j\backslash j}\mathbf{h}_{jlk}^H+\tfrac{1}{\sqrt{\rho_{\rm{tr}}}}\mathbf{n}_{jk}^H\right) \mathbf{Q}\mathbf{R}_{jj}\mathbf{\Pi}_{jk}\mathbf{h}_{jjk}
\nonumber\\
&+\tfrac{2}{M^2}\mathfrak{Re}\left\{\mathbf{h}_{jjk}^H\mathbf{\Pi}_{jk}\mathbf{R}_{jj}\mathbf{Q} \left(\sum\nolimits_{l\in \mathcal{S}_j\backslash j}\mathbf{h}_{jlk}+\tfrac{1}{\sqrt{\rho_{\rm{tr}}}}\mathbf{n}_{jk}\right) \mathbf{h}_{jjk}^H\mathbf{Q}\mathbf{R}_{jj}\mathbf{\Pi}_{jk}\mathbf{h}_{jjk}\right\} \nonumber,
\end{align}
where the third term on the right-hand side converges to zero almost surely by Lemma 4 (iii) in \cite{HowManyAntennas13}. By using similar way to derive \eqref{E:Asymp_3}, we can show that the first term converges to $|\tfrac{1}{M}\mathbf{h}_{jjk}^H\mathbf{\Pi}_{jk}\mathbf{R}_{jj}\mathbf{Q}\mathbf{h}_{jjk}|^2 \!\xrightarrow[M,K\rightarrow \infty]{a.s.}\tfrac{1}{M^2}\delta_{j}(\varphi)^2$.
By using Lemma 4 (ii) in \cite{HowManyAntennas13} twice and noting that $\mathbf{\Phi}_{jj}=v\mathbf{R}_{jj}$, the second term almost surely converges to $\frac{(1-v)}{M^4}\mathrm{tr}(\mathbf{R}_{jj} \mathbf{T}_{j}^{''}(\varphi))=\frac{(1-v)}{vM^3}\delta_j^{''}(\varphi)$, where $\mathbf{T}_{j}^{''}(\varphi)=\mathbf{T}_{j}(\varphi)\left(\mathbf{\Phi}_{jj}+\frac{K}{M} \frac{\mathbf{\Phi}_{jj}\delta_{j}^{''}(\varphi)}{(1+\delta_{j}(\varphi))^2}\right)\mathbf{T}_{j}(\varphi)$, and $\delta_{j}^{''}(\varphi)=\frac{1}{M}\mathrm{tr}(\mathbf{\Phi}_{jj}\mathbf{T}_{j}^{''}(\varphi))$.
Define $\mathbf{T}_{j}^{''0}=\lim_{\varphi\rightarrow0}\varphi^2 \mathbf{T}_{j}^{''}(\varphi)=\left(1+ \frac{K\delta_{j}^{''0}}{M(\delta^{0})^2}\right)\mathbf{T}^0 \mathbf{\Phi}_{jj}\mathbf{T}^0$, and $\delta_{j}^{''0}=\frac{1}{M}\mathrm{tr}(\mathbf{\Phi}_{jj}\mathbf{T}_{j}^{''0})$. Upon substituting $\mathbf{\Phi}_{jj}=v\alpha\rho \tilde{\mathbf{R}}\tilde{\mathbf{R}}^H$ and $\mathbf{T}^0=(\frac{\rho K}{M-\rho K}\tilde{\mathbf{R}}\tilde{\mathbf{R}}^H+\mathbf{I})^{-1}$, we can derive that $\delta_j^{''0}=v\alpha\rho \delta^{0}$.

Based on the above results, we can obtain the signal power~as
\begin{align}\label{E:Asymp_Sig_ZF}
\bar{\lambda}_{jk}|\tfrac{1}{M}\mathbf{h}_{jjk}\mathbf{\Pi}_{j}\mathbf{\hat{h}}_{jjk}|^2 &\xrightarrow[M,K\rightarrow \infty]{a.s.}\frac{1}{M}\frac{M(\delta_{j}(\varphi))^2 +(\frac{1}{v}-1)\delta_{j}^{''}(\varphi)} {\delta_{j}^{'}(\varphi)} \nonumber \\& \xrightarrow[\varphi\rightarrow 0]{}\frac{v\alpha(M+\tfrac{1}{\gamma_{\rm{tr}}\alpha}+\rho (\bar{L}_P-1-K))}{M}.
\end{align}

\subsubsection{Coherent ICI Caused by PC} By using the same way to derive the signal power and considering that $\mathbf{\Phi}_{lj}=\mathbf{R}_{ll}\mathbf{Q}\mathbf{R}_{lj}=\chi \mathbf{\Phi}_{ll}$, we can show that the coherent ICI caused by PC converges almost surely to
\begin{align}\label{E:Asymp_ICI_PC_ZF}
\bar{\lambda}_{lk}\left|\tfrac{1}{M}\mathbf{h}_{ljk}^H\mathbf{\Pi}_{l}\mathbf{\hat{h}}_{llk}\right|^2
&\xrightarrow[M,K\rightarrow \infty]{a.s.} \frac{1}{M} \frac{M(\chi\delta_l(\varphi))^2+(\frac{1}{v}-\chi)\chi\delta_{l}^{''}(\varphi)} {\delta_{l}^{'}(\varphi)}\nonumber\\
&\xrightarrow[\varphi\rightarrow 0]{}\frac{\chi v\alpha(M\chi+ \frac{1}{\alpha\gamma_{\rm{tr}}}+\rho(\bar{L}_P-\chi-\chi K))}{M}.
\end{align}

\subsubsection{MUI and Non-coherent ICI} According to the proof of Theorem 6 in \cite{HowManyAntennas13}, the interference caused by $l\in \mathcal{S}_j$ and $m\neq k$ of RZF~is
\begin{align}
\bar{\lambda}_{lm}|\tfrac{1}{M}\mathbf{h}_{ljk}^H\mathbf{\Pi}_{l}\mathbf{\hat{h}}_{llm}|^2 \xrightarrow[M,K\rightarrow \infty]{a.s.} & \tfrac{1}{M^2}\frac{\mathrm{tr}(\mathbf{R}_{lj}\mathbf{T}_{l}^{''}(\varphi))}{\delta_{l}^{'}(\varphi)}
+ \tfrac{1}{M^3} \frac{|\mathrm{tr}(\mathbf{\Phi}_{lj}\mathbf{T}_l(\varphi))|^2 \delta_{l}^{''}(\varphi)} {(1+\delta_{l}(\varphi))^2\delta_{l}^{'}(\varphi)} \\& - \tfrac{2}{M^3}\mathfrak{Re}\left\{ \frac{\mathrm{tr}(\mathbf{\Phi}_{lj}\mathbf{T}_{l}^{''}(\varphi)) \left(\mathrm{tr}(\mathbf{\Phi}_{lj}\mathbf{T}_l(\varphi))\right)^{*}} {(1+\delta_{l}(\varphi))\delta_{l}^{'}(\varphi)}\right\}\nonumber.
\end{align}

For $l=j$ and $m\neq k$, when $\varphi\rightarrow 0$, the MUI of ZFBF converges almost surely to
\begin{equation}\label{E:Asymp_MUI_noPC_ZF}
\bar{\lambda}_{jm}|\tfrac{1}{M}\mathbf{h}_{jjk}^H\mathbf{\Pi}_{j}\mathbf{\hat{h}}_{jjm}|^2 \xrightarrow[M,K\rightarrow \infty, \varphi \rightarrow 0]{a.s.} \frac{(1-v)\alpha\rho}{M}.
\end{equation}

For $l\in \mathcal{S}_j\backslash j$ and $m\neq k$, when $\varphi\rightarrow 0$, the non-coherent ICI of ZFBF caused by the users in the cells using the same set of pilot sequences converges almost surely to
\begin{equation}\label{E:Asymp_ICI_noPC_ZF2}
\bar{\lambda}_{lm}|\tfrac{1}{M}\mathbf{h}_{ljk}^H\mathbf{\Pi}_{l}\mathbf{\hat{h}}_{llm}|^2 \xrightarrow[M,K\rightarrow \infty, \varphi \rightarrow 0]{a.s.} \frac{(1-\chi v)\chi \alpha\rho}{M}.
\end{equation}

For $l\notin \mathcal{S}_j$ and arbitrary $m$, the non-coherent ICI of ZFBF caused by the users in the cells using different set of pilot sequences can be derived as
\begin{equation}\label{E:Asymp_ICI_noPC_ZF1}
\bar{\lambda}_{lm}|\tfrac{1}{M}\mathbf{h}_{ljk}^H\mathbf{\Pi}_{l}\mathbf{\hat{h}}_{llm}|^2 \xrightarrow[M,K\rightarrow \infty]{a.s.}\frac{\chi}{vM}\frac{\delta_l^{''}(\varphi)} {\delta_{l}^{'}(\varphi)}\xrightarrow[\varphi\rightarrow0]{} \frac{\chi \alpha\rho}{M}.
\end{equation}

By substituting \eqref{E:Asymp_Sig_ZF}, \eqref{E:Asymp_ICI_PC_ZF}, \eqref{E:Asymp_MUI_noPC_ZF}, \eqref{E:Asymp_ICI_noPC_ZF2} and \eqref{E:Asymp_ICI_noPC_ZF1} into \eqref{E:SINR_RZF}, we can obtain the asymptotic SINR with ZFBF~as
\begin{align*}
&\gamma_{jk}^{\rm{Z}}\xrightarrow[M,K\rightarrow\infty]{a.s.} \bar{\gamma}_{jk}^{\rm{Z}}=\frac{M+\frac{1}{\gamma_{\rm{tr}}\alpha} +\rho(\bar{L}_P-1-K)} {\chi(\bar{L}_P-1)(M-2\rho K)
+(K\bar{L}-1)(\rho\bar{L}_P+\frac{1}{\gamma_{\rm{tr}}\alpha}) -\rho(K\!-\!1)\!+\!\frac{K\sigma^2}{v\alpha P}},
\end{align*}
where $\bar{L}_P=1+\chi(L_P-1)$, $\bar{L}=1+\chi(L-1)$ and $v=\frac{\alpha\rho}{\alpha\rho\bar{L}_P+\frac{1}{\gamma_{\rm{tr}}}}$.

Then, the asymptotic data rate per BS with ZFBF can be obtained as
\begin{equation}
\bar{R}_{\rm{BS}}^{\rm{Z}}=\lim_{M,K\rightarrow \infty} R_j^{\rm{Z}} = BK\log_2\left(1+\bar{\gamma}_{jk}^{\rm{Z}}\right).
\end{equation}
%
%\section{Proof of Proposition 1} \label{P:Proof_P1}
%According to \eqref{E:KKT_Accurate}, we can express the $MP_0$ as a function of $P^*$ as
%\begin{equation}
%\frac{MP_0}{(1-\frac{T_{\rm{tr}}}{T})\eta}= \frac{((S+I)P^*+G)(IP^*+G)}{SG}\ln\left(1+\frac{SP^{*}}{IP^{*}+G}\right)-P^*.
%\end{equation}
%The first derivative can be derived as
%\begin{equation}
%\frac{d\left(\tfrac{MP_0}{(1-\frac{T_{\rm{tr}}}{T})\eta}\right)}{dP^*} = \frac{2(S+I)IP^*+(S+2I)G}{SG}\ln\left(1+\frac{SP^{*}}{IP^{*}+G}\right)>0.
%\end{equation}
%It is obvious that $\frac{dP^*}{d\left(\tfrac{MP_0}{(1-\frac{T_{\rm{tr}}}{T})\eta}\right)} = \frac{1}{\frac{d\left(\tfrac{MP_0}{(1-\frac{T_{\rm{tr}}}{T})\eta}\right)}{dP^*}}>0$, which indicates that $P^*$ increases monotonically with $\frac{MP_0}{(1-\frac{T_{\rm{tr}}}{T})\eta}$.
%

\section{Approximated Optimal Transmit Power $P^*$}\label{P:Optimal_P_MRT}
\renewcommand{\theequation}{B.\arabic{equation}}
The KKT condition of $P^*$ in \eqref{E:KKT_Accurate} can be expressed~as
\setcounter{equation}{0}
\begin{align}
&\frac{SG(P^*+\frac{MP_0}{(1-\frac{KT_{\rm{tr}}}{T})\eta})} {((S+I)P^*+G)(IP^*+G)} -\ln\left(1+\frac{SP^*}{IP^*+G}\right)\nonumber\\
&=\frac{\frac{MP_0}{(1-\frac{KT_{\rm{tr}}}{T})\eta}- \frac{G}{S+I}}{P^*+\frac{G}{S+I}} \!-\!\frac{\frac{MP_0}{(1-\frac{KT_{\rm{tr}}}{T})\eta}\! -\!\frac{G}{I}}{P^*+\frac{G}{I}} \!-\!\ln\left(\!1\!+\!\frac{G}{(S+I)P^*}\!\right) \!+\!\ln\left(\!1\!+\!\frac{G}{IP^*}\!\right) \!-\!\ln\left(\!1\!+\!\frac{S}{I}\!\right)\nonumber\\
&\mathop\approx^{(a)} \frac{\frac{MP_0}{(1-\frac{KT_{\rm{tr}}}{T})\eta}- \frac{G}{S+I}}{P^*+\frac{G}{S+I}} \!-\!\frac{\frac{MP_0}{(1-\frac{KT_{\rm{tr}}}{T})\eta}\!- \!\frac{G}{I}}{P^*+\frac{G}{I}} \!-\!\frac{\frac{G}{(S+I)}}{P^*+\frac{G}{2(S+I)}} \!+\!\frac{\frac{G}{I}}{P^*+\frac{G}{2I}} \!-\!\ln\left(\!1\!+\!\frac{S}{I}\!\right)\nonumber\\
&\mathop\approx^{(b)} \frac{\frac{MP_0}{(1-\frac{KT_{\rm{tr}}}{T})\eta}- \frac{2G}{S+I}} {P^*+\frac{G}{S+I}} \!-\!\frac{\frac{MP_0}{(1-\frac{KT_{\rm{tr}}}{T})\eta}\! -\!\frac{2G}{I}}{P^*+\frac{G}{I}} \!-\!\ln\left(\!1\!+\!\frac{S}{I}\!\right)\nonumber\\
&\mathop\approx^{(c)} \frac{\frac{MP_0}{(1-\frac{KT_{\rm{tr}}}{T})\eta}}{P^*+\frac{G}{S+I}} \!-\!\frac{\frac{MP_0}{(1-\frac{KT_{\rm{tr}}}{T})\eta}}{P^*+\frac{G}{I}} \!-\!\ln\left(\!1\!+\!\frac{S}{I}\!\right)=0,\label{E:KKT_Approx}
\end{align}
where approximation (a) comes from the first order approximation of $\ln(1+x)\approx \frac{2x}{2+x}$, which is accurate for small $x$. When $M$ is very large, both $S$ and $P^*$ are large according to Proposition 1. Furthermore, if the system is pilot contaminated, $I$ increases with $M$ and is also very large. Therefore, approximation (a) is accurate for large value of $M$. Approximation (b) comes from $P^*+\frac{G}{2(S+I)}\approx P^*+\frac{G}{(S+I)}$ and $P^*+\frac{G}{2 I}\approx P^*+\frac{G}{I}$, since $P^*\gg \frac{G}{S+I}$ and $P^*\gg \frac{G}{I}$ for large $P^*$ and $S$, which are true for large value of $M$. Approximation (c) comes from $\frac{MP_0}{(1-\frac{KT_{\rm{tr}}}{T})\eta}\gg \frac{G}{I}>\frac{G}{S+I}$ because of large values of $M$ and $S$ in massive MIMO. Furthermore, if $P_0$ is large, the value of $P^*$ will be large, then the approximations are more accurate.

The approximate solution of \eqref{E:KKT_Accurate} (i.e., the solution of \eqref{E:KKT_Approx})~is
\begin{align}
P^*&\approx\left(\frac{G}{I}-\frac{G}{S+I}\right) \sqrt{\frac{1}{4}+\frac{\frac{MP_0}{(1-\frac{KT_{\rm{tr}}}{T})\eta}} {(\frac{G}{I}-\frac{G}{S+I}) \ln(1+\frac{S}{I})}} -\frac{1}{2}\left(\frac{G}{I}+\frac{G}{S+I}\right)\nonumber\\
&\mathop\approx^{(d)} \sqrt{\frac{\frac{MP_0}{(1-\frac{KT_{\rm{tr}}}{T})\eta}(\frac{G}{I} -\frac{G}{S+I})}{\ln(1+\frac{S}{I})}} -\frac{1}{2}\left(\frac{G}{I}+\frac{G}{S+I}\right)
\mathop\approx^{(e)} \sqrt{\frac{MP_0K\sigma^2}{(1-\frac{KT_{\rm{tr}}}{T})\eta v}} \sqrt{\frac{\frac{1}{I}-\frac{1}{S+I}}{\ln(1+\frac{S}{I})}},
\end{align}
where approximation (d) comes from ignoring $\frac{1}{4}$ because $\frac{\frac{MP_0}{(1-\frac{KT_{\rm{tr}}}{T})\eta}} {(\frac{G}{I}-\frac{G}{S+I}) \ln(1+\frac{S}{I})}\gg \frac{1}{4}$ for large values of $S$ and $MP_0$ due to large $M$, and approximation (e) comes from ignoring the second term $\frac{1}{2}\left(\frac{G}{I}+\frac{G}{S+I}\right)$ because the first term (proportional to $\frac{1}{\sqrt{\ln M}}$) is much larger than the second term (proportional to $\frac{1}{M}$).

According to the analysis above, approximations (a)-(e) are all accurate when the value of $M$ is large and become even more accurate when the value of $P_0$ is large.
\bibliography{IEEEabrv,Liubib}

\end{document}